\documentclass[twocolumn,showpacs,showkeys,preprintnumbers,amssymb,aps,superscriptaddress,prb]{revtex4-2}
\usepackage{graphicx}
\usepackage{dcolumn}
\usepackage{bm}
\usepackage{amssymb}
\usepackage{epsfig}
\usepackage{color}

\begin{document}

\title{Frustrated magnetism of spin-1/2 Heisenberg diamond and octahedral chains as a statistical-mechanical monomer-dimer problem}
\author{Jozef Stre\v{c}ka}
\email{jozef.strecka@upjs.sk}
\affiliation{Department of Theoretical Physics and Astrophysics, Faculty of Science, \\
P. J. \v{S}af\'{a}rik University, Park Angelinum 9, 04001 Ko\v{s}ice, Slovakia}
\author{Taras Verkholyak}
\affiliation{Institute for Condensed Matter Physics, NASU, Svientsitskii Street 1, 79011 L'viv, Ukraine}
\author{Johannes Richter} 
\affiliation{Institut f\"ur Theoretische Physik, Otto-von-Guericke Universit\"at in Magdeburg, 39016 Magdeburg, Germany}
\affiliation{Max-Planck-Institut f\"ur Physik Komplexer Systeme, 01187 Dresden, Germany}
\author{Katar\'ina Karl'ov\'a} 
\affiliation{Department of Theoretical Physics and Astrophysics, Faculty of Science, \\
P. J. \v{S}af\'{a}rik University, Park Angelinum 9, 04001 Ko\v{s}ice, Slovakia}
\author{Oleg Derzhko} 
\affiliation{Institute for Condensed Matter Physics, NASU, Svientsitskii Street 1, 79011 L'viv, Ukraine}
\author{J\"urgen Schnack} 
\affiliation{Department of Physics, Bielefeld University, P.O. Box 100131, D-33501 Bielefeld, Germany}

\date{\today}

\begin{abstract}
It is evidenced that effective lattice-gas models of hard-core monomers and dimers afford a proper description of low-temperature features of spin-1/2 Heisenberg diamond and octahedral chains. Besides monomeric particles assigned within the localized-magnon theory to bound one- and two-magnon eigenstates, the effective monomer-dimer lattice-gas model additionally includes dimeric particles assigned to a singlet-tetramer (singlet-hexamer) state as a cornerstone of dimer-tetramer (tetramer-hexamer) ground state of a spin-1/2 Heisenberg diamond (octahedral) chain. A feasibility of the effective description is confirmed through the exact diagonalization and finite-temperature Lanczos methods. Both quantum spin chains display rich ground-state phase diagrams including discontinuous as well as continuous field-driven phase transitions, whereby the specific heat shows in vicinity of the former phase transitions an extraordinary low-temperature peak coming from a highly-degenerate manifold of low-lying excitations.  
\end{abstract}


\maketitle

\section{Introduction}
\label{sec:intro}

Frustrated quantum Heisenberg spin systems remain for a long time at the forefront of research interest, because they often exhibit intriguing ordered or disordered quantum ground states in addition to unconventional magnetic properties emergent at sufficiently low temperatures \cite{lacr10,diep20}. From this perspective, it is quite bizarre that a more complex nature of exchange pathways initiating geometric spin frustration simultaneously offers a genuine possibility of finding quantum ground states of frustrated Heisenberg spin models by the use of rigorous methods. Among a few highly celebrated examples of frustrated quantum Heisenberg spin models with exactly known ground states one could for instance mention the zig-zag ladder \cite{maju69}, the Shastry-Sutherland lattice \cite{shas81}, the delta chain \cite{mont91}, and the one-dimensional chain of linked tetrahedra \cite{gelf91}. 

Compared to this, it is generally much more difficult to cope magnetic and thermodynamic properties of frustrated quantum Heisenberg spin models at low but non-zero temperatures. The unbiased exact diagonalization (ED) \cite{scna10,snac10} or finite-temperature Lanczos method (FTLM) \cite{schn10,hane14,schm17,prel18,schn18,schn20} are regrettably limited to relatively small finite-size Heisenberg spin systems involving at most a few dozens of quantum spins. On the other hand, the large-scale simulations of frustrated Heisenberg spin models based on quantum Monte Carlo (QMC) methods suffer from a serious negative-sign problem when performed in a standard basis, and one should accordingly resort to more challenging sign-problem-free QMC computations performed in a dimer \cite{hone16,wess17,wess18,stap18,demi20} or trimer \cite{webe21} basis.

A flat band emergent in the one-magnon spectrum of frustrated Heisenberg spin systems bears evidence of a magnon, which is bound to a small portion of the frustrated spin lattice due to a destructive quantum interference. Owing to this fact, the many-magnon eigenstates of frustrated Heisenberg spin models can be simply designed from a bound one-magnon eigenstate within the concept of localized magnons establishing an effective lattice-gas model \cite{schu02}, which consists in an independent arrangement of bound magnons on trapping cells of frustrated spin lattices \cite{zhit05,derz06,derz15}. 

It should be pointed out, however, that the effective lattice-gas models developed within the standard formulation of the localized-magnon theory are usually eligible for a description of frustrated quantum Heisenberg spin systems only in a high-field region sufficiently close to the saturation field \cite{zhit05,derz06,derz15}. One may fortunately avoid this shortcoming in a special subclass of the frustrated quantum Heisenberg spin models satisfying local spin-conservation laws. It has been actually verified that an additional counting of the lowest-energy bound two-magnon eigenstate paves the way towards a complete description of low-temperature magnetization curves and thermodynamics of highly frustrated Heisenberg octahedral chains from zero up to the saturation field \cite{stre17,stre18,karl19,stre20}. 

In the present paper we will generalize the localized-magnon theories established previously for the spin-1/2 Heisenberg diamond and octahedral chains by extending their validity to a less frustrated parameter space including dimer-tetramer \cite{taka96} and tetramer-hexamer \cite{stre17} ground states, respectively. It is worthwhile to remark that the spin-1/2 Heisenberg diamond chain bears a close relation to several copper-based magnetic compounds such as azurite Cu$_3$(CO$_3$)$_2$(OH)$_2$ \cite{jesc11,hone11}, A$_3$Cu$_3$AlO$_2$(SO$_4$)$_4$ (A = K, Rb, and Cs) \cite{fuji15,mori17,fuji17},  
[Cu$_3$(OH)$_2$(CH$_3$COO)(H$_2$O)$_4$](RSO$_3$)$_2$ (RSO$_3$ = organic sulfonate anions) \cite{fuji16} and Cu$_3$ (CH$_3$COO)$_4$(OH)$_2$ $\cdot$ 5H$_2$O \cite{cuim19}. Although we do not have at present any knowledge about a possible experimental realization of the spin-1/2 Heisenberg octahedral chain a relatively numerous family of  hexanuclear complexes with the magnetic core of a discrete copper-based octahedron \{Cu$_6$\} \cite{liuc03,xian05,zhao15,gaox16} could be used for a brick-and-mortar synthesis of a one-dimensional copper polymeric chain of corner-sharing octahedra. A closely related extended magnetic structure with the architecture of a three-dimensional network of corner-sharing octahedra \{Cu$_6$\} is for instance realized in the copper-based metal-organic framework [Cu$_3$(tpt)$_4$](ClO$_4$)$_3$ \cite{robs96}. Last but not least, an artificial design of the spin-1/2 Heisenberg octahedral chain achieved through magnetic atom-trap lattices is also feasible with regard to a recent successful realization of several related one-dimensional magnetic structures such as ladders and diamond spin chains \cite{rooi19}. 

The outline of this paper is as follows. The overall ground-state phase diagrams of the spin-1/2 Heisenberg diamond and octahedral chains will be presented in Sec.~\ref{sec:dc} and \ref{sec:oc} together with a few basic steps elucidating its construction from exact and density-matrix renormalization group (DMRG) calculations of the effective mixed-spin Heisenberg chains. The spin-1/2 Heisenberg diamond and octahedral chains will be reformulated within the modified localized-magnon theory establishing their connection to a one-dimensional lattice-gas model of hard-core monomers and dimers in Sec.~\ref{sec:dcmd} and \ref{sec:ocmd}, where an eligibility of the effective description will be also corroborated through a detailed comparison with numerical data obtained from the full ED and FTLM. The paper ends up with a brief summary of the most important findings in Sec.~\ref{sec:con}, where a few future outlooks are also presented.

\section{Ground states of a spin-1/2 Heisenberg diamond chain}
\label{sec:dc}
\begin{figure}
\centering\includegraphics[width=0.9\columnwidth]{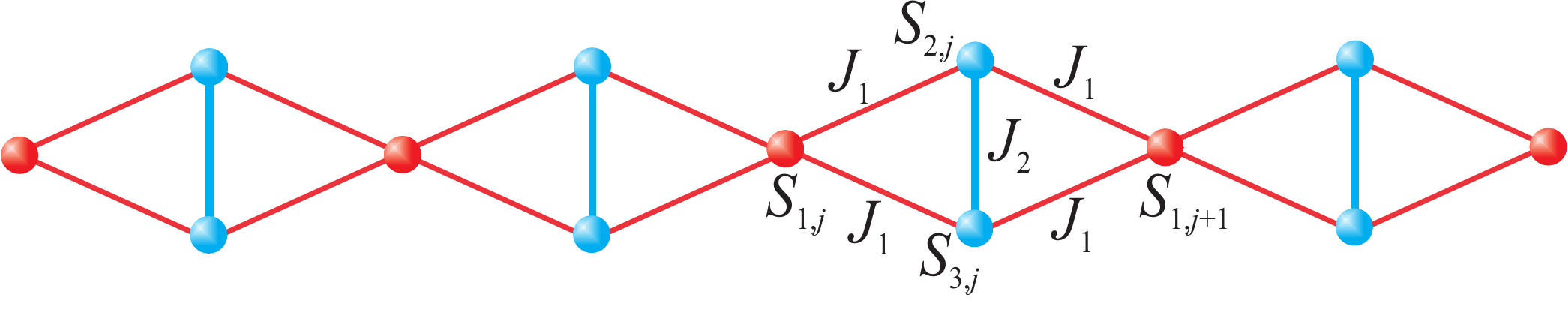} 
\vspace*{-0.4cm}
\caption{A schematic illustration of the spin-1/2 Heisenberg diamond chain including the notation for the lattice sites and assumed coupling constants.}
\label{fig:dc}
\end{figure}

First, let us consider the spin-1/2 Heisenberg diamond chain, which is schematically illustrated in Fig. \ref{fig:dc} and mathematically defined through the Hamiltonian
\begin{eqnarray}
\hat{\cal H} = 
\sum_{j=1}^{N} \Bigl[ J_1 (\boldsymbol{\hat{S}}_{1,j} &+& \boldsymbol{\hat{S}}_{1,j+1}) \!\cdot\! (\boldsymbol{\hat{S}}_{2,j} + \boldsymbol{\hat{S}}_{3,j}) \nonumber \\ 
+ J_2 \boldsymbol{\hat{S}}_{2,j}\!\cdot\!\boldsymbol{\hat{S}}_{3,j} &-& h \sum_{i=1}^{3} \hat{S}_{i,j}^{z} \Bigr], 
\label{hamdc}
\end{eqnarray}
where $\hat{\bf{S}}_{i,j}$ $(i=1,2,3; j=1,\ldots,N)$ denotes spin-1/2 operator assigned to a lattice site unambiguously given by two subscripts, the first subscript specifies spins from a given unit cell and the second subscript  determines the unit cell itself. The coupling constant $J_1>0$ stands for the antiferromagnetic interaction between nearest-neighbor monomeric (${S}_{1,j}$) and dimeric (${S}_{2,j}$, ${S}_{3,j}$) spins, while the coupling constant $J_2>0$ accounts for the antiferromagnetic intradimer interaction being responsible for a geometric spin frustration. The Zeeman term $h \geq 0$ accounts for an effect of the external magnetic field and finally, the periodic boundary condition is imposed for simplicity ${S}_{1,N}\equiv {S}_{1,1}$.

\begin{figure}
\centering\includegraphics[width=0.9\columnwidth]{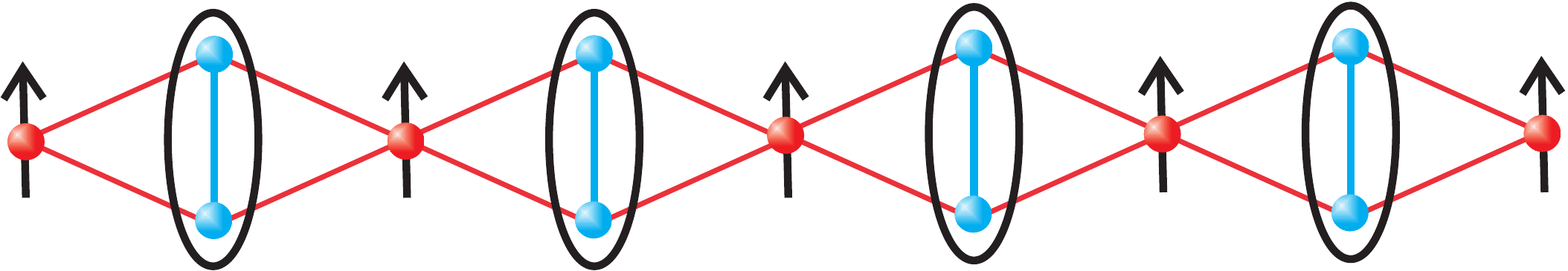} 
\vspace*{-0.4cm}
\caption{A schematic illustration of the monomer-dimer ground state (\ref{vecMD}), which is responsible for occurrence of intermediate one-third plateau in a zero-temperature magnetization curve. An oval denotes a singlet-dimer state.}
\label{fig:dc2}
\end{figure}

The ground-state phase diagram of the spin-1/2 Heisenberg diamond chain given by Hamiltonian (\ref{hamdc}) can be obtained by combining three complementary techniques. One may adapt variational arguments (see Appendix in Ref. \cite{verk11}) in order to rigorously prove in the highly frustrated parameter region $J_2>2J_1$ and low enough magnetic fields $h<J_1+J_2$ the monomer-dimer (MD) ground state schematically illustrated in Fig. \ref{fig:dc2} and mathematically given by the eigenvector   
\begin{eqnarray}
|{\rm MD} \rangle \!=\! \left\{ \begin{array}{l}
\! \displaystyle \prod_{j=1}^N  |{S}_{1,j}\rangle \!\otimes\! \frac{1}{\sqrt{2}}(|\!\uparrow_{2,j}\downarrow_{3,j}\rangle \!-\! |\!\downarrow_{2,j}\uparrow_{3,j}\rangle), \,\,\, h = 0,\\
\! \displaystyle \prod_{j=1}^N |\!\uparrow_{1,j}\rangle \!\otimes\! 
\frac{1}{\sqrt{2}}(|\!\uparrow_{2,j}\downarrow_{3,j}\rangle \!-\! |\!\downarrow_{2,j}\uparrow_{3,j}\rangle), 
\,\,\, h > 0,
\end{array} \right. 
\label{vecMD}
\end{eqnarray}
where $|{S}_{1,j}\rangle$ denotes any out of two available states $|\!\!\uparrow_{1,j}\rangle$ and $|\!\!\downarrow_{1,j}\rangle$ of the monomeric spins. It is worthwhile to remark that the MD ground state (\ref{vecMD}) is macroscopically degenerate at zero magnetic field due to frustrated (paramagnetic) character of all monomeric spins ${S}_{1,j}$, which are consequently perfectly aligned into the magnetic-field direction once the external field is turned on. Another independent confirmation of the MD ground state (\ref{vecMD}) is provided by the localized-magnon approach, which restricts its presence in the highly frustrated parameter space $J_2>2J_1$ to sufficiently low magnetic fields $h< h_s$ not exceeding the saturation field $h_s = J_1+J_2$ associated with a field-driven phase transition to the fully saturated ferromagnetic phase \cite{derz06,derz07,derz12}. It could be thus concluded that the variational and localized-magnon theories give evidence of the MD ground state (\ref{vecMD})  from zero up to the saturation field in the highly frustrated parameter region $J_2>2J_1$. 

\begin{figure}
\centering\includegraphics[width=0.9\columnwidth]{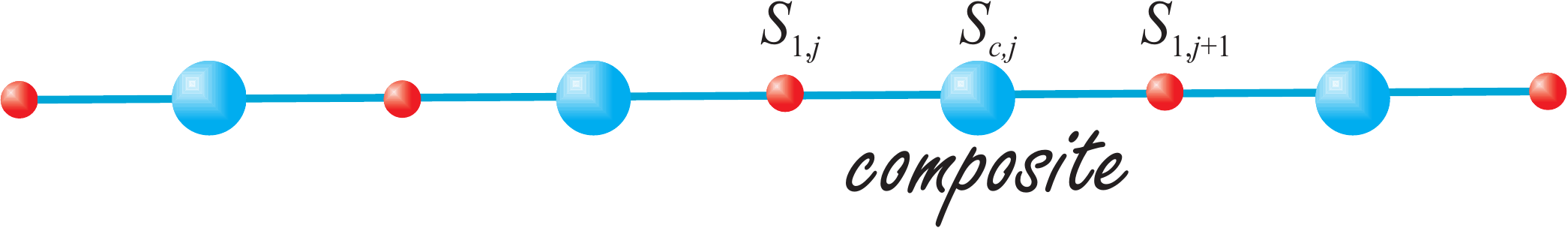} 
\vspace{-0.4cm}
\caption{A schematic illustration of the effective mixed spin-(1/2, $S_{c,j}$) Heisenberg chains, in which the composite spins may acquire the values $S_{c,j}=0$ or 1 for a diamond chain and $S_{c,j}=0$, 1 or 2 for an octahedral chain.}
\label{fig:eff}
\end{figure}

In the rest of the parameter space $J_2<2J_1$, the ground states of the spin-1/2 Heisenberg diamond chain cannot be simply found on the grounds of exact analytical calculations. To this end, it is convenient to rewrite the zero-field part of the Hamiltonian (\ref{hamdc}) into the following equivalent form
\begin{eqnarray}
\hat{\cal H} = J_1 \sum_{j=1}^{N} (\boldsymbol{\hat{S}}_{1,j} \!+\! \boldsymbol{\hat{S}}_{1,j+1}) \!\cdot\! \boldsymbol{\hat{S}}_{c, j} 
+ \frac{J_2}{2} \sum_{j=1}^{N} \!\left(\boldsymbol{\hat{S}}_{c, j}^2 \!-\! \frac{3}{2}\right)\!\!,
\label{effms}
\end{eqnarray}
which directly implies a local conservation of the composite spin $\hat{\bf{S}}_{c,j}=\hat{\bf{S}}_{2,j} + \hat{\bf{S}}_{3,j}$ on vertical dimers of a spin-1/2 Heisenberg diamond chain. The composite spin on vertical dimers is accordingly conserved quantity with well defined quantum spin numbers $S_{c,j} = 0$ or 1. Hence, it follows that all ground states of the spin-1/2 Heisenberg diamond chain can be derived from the lowest-energy eigenstates of effective mixed spin-($1/2, S_{c,j}$) Heisenberg chains schematically illustrated in Fig. \ref{fig:eff} when performing appropriate shift of their eigenenergies according to the second term of the Hamiltonian (\ref{effms}). Note furthermore that the $z$-component of the total spin operator $\hat{S}_T^z = \sum_{j=1}^{N} \sum_{i=1}^{3} \hat{S}_{i,j}^{z}$ commutes with the Hamiltonian (\ref{hamdc}), which means that most preferable energy eigenvalues of the spin-1/2 Heisenberg diamond chain in a nonzero magnetic field directly follow from their zero-field counterparts when performing just a trivial shift by the respective Zeeman's term
\begin{eqnarray}
E(S_T^z, h \neq 0) = E(S_T^z, h=0) - h S_T^z.
\end{eqnarray}

\begin{figure}
\centering\includegraphics[width=0.9\columnwidth]{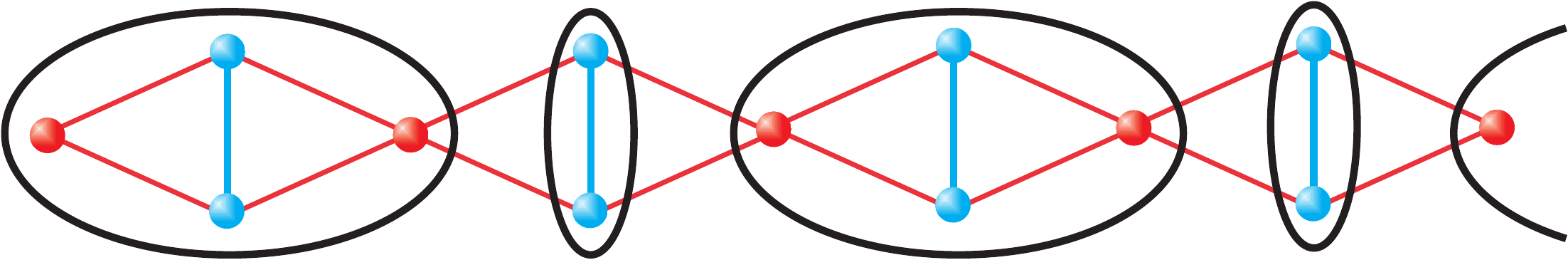}
\vspace{-0.4cm}
\caption{A schematic illustration of the singlet tetramer-dimer phase given by the eigenvector (\ref{tdgs}). Large and small ovals represent regularly alternating singlet-tetramer and singlet-dimer states.}
\label{fig:dctd}
\end{figure}

\begin{figure}
\centering\includegraphics[width=\columnwidth]{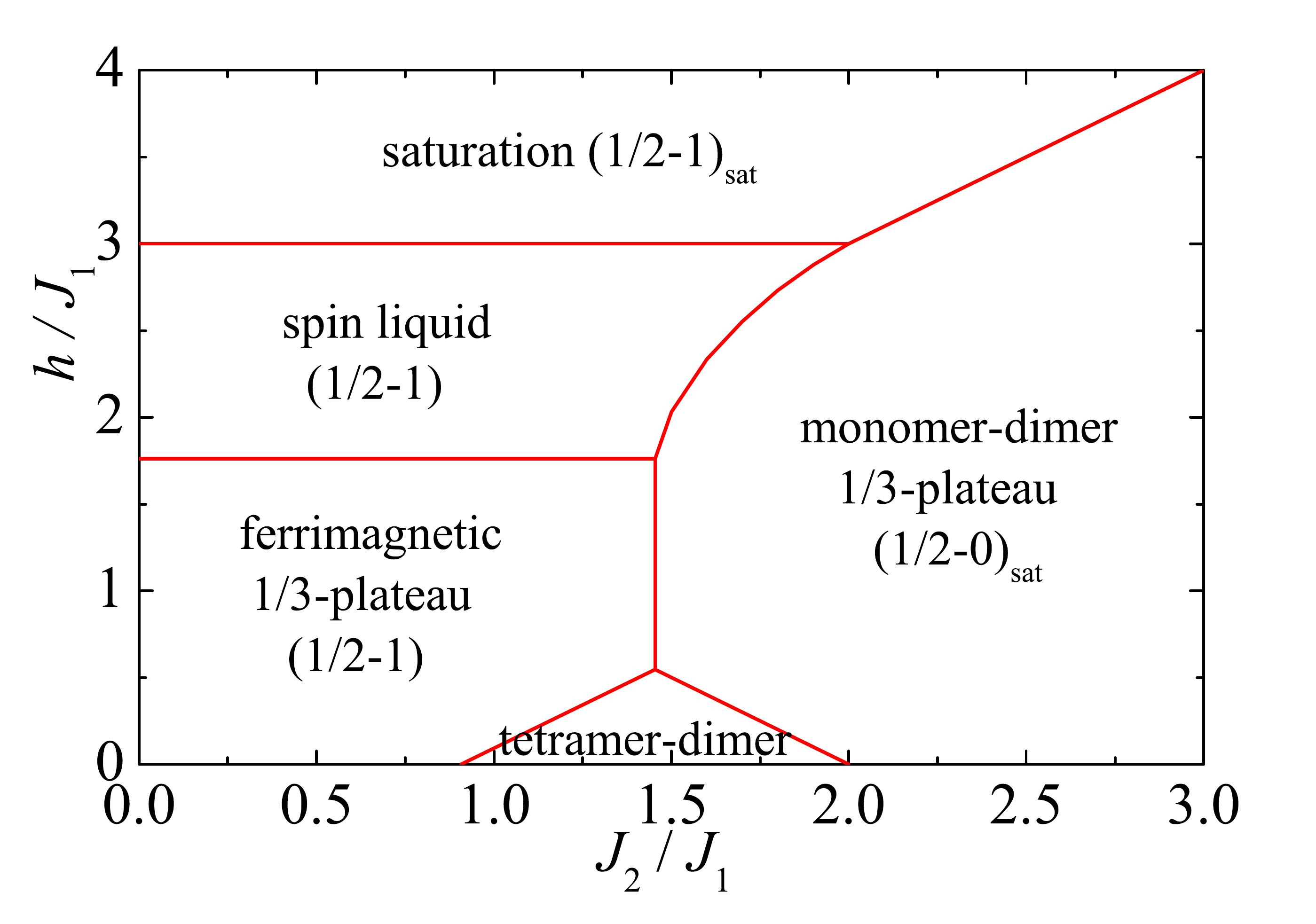}
\vspace{-0.9cm}
\caption{The ground-state phase diagram of the spin-1/2 Heisenberg diamond chain in $J_2/J_1-h/J$ plane. The numbers in parentheses determine spin values within the effective mixed spin-(1/2, $S_c$) Heisenberg chain.}
\label{fig:gsdc}
\end{figure}

Although one should generally find out lowest-energy eigenstates of effective mixed spin-($1/2, S_{c,j}$) Heisenberg chains (\ref{effms}) with all possible combinations of the composite spins ($\forall j$ $S_{c,j} = 0$ or $1$) it is often sufficient to consider just a few particular cases with quantum spin numbers that do not break translational symmetry or at most limit the spontaneous breaking of translational symmetry to a few unit cells. By inspection we have found that all ground states of the spin-1/2 Heisenberg diamond chain stem from the effective mixed spin-($1/2, 0$) or mixed spin-($1/2, 1$)  Heisenberg chains without translationally broken symmetry or the effective mixed spin-(1/2, 1, 1/2, 0) Heisenberg chain with a period doubling. The lowest-energy eigenstate of the effective mixed spin-($1/2, 0$) Heisenberg chain with the unique choice of zero composite spin $\forall j$ $S_{c,j} = 0$ naturally corresponds to the MD ground state (\ref{vecMD}) with the exact energy
\begin{eqnarray}
E_{1/2-0} (N, S_T^z = N/2) = - \frac{3}{4} N J_2 - \frac{1}{2} N h,
\label{eff120}
\end{eqnarray}
because the singlet-dimer state at vertical bonds assigned to zero composite spin decouples all correlations between the dimeric and monomeric spins. It is worthwhile to remark, moreover, that the effective mixed-spin Heisenberg chains with regularly alternating zero composite spins are always fragmented into smaller quantum spin clusters, for which the lowest-energy eigenstates can be rather easily calculated by the ED method. On the other hand, the fragmentation does not occur for the other unique choice of the composite spins $\forall j$ $S_{c,j} = 1$, which contrarily gives rise to highly correlated collective eigenstates with the energy eigenvalues governed by the formula
\begin{eqnarray}
E_{1/2-1} (N, S_T^z) \!=\! N J_1 \varepsilon_{1/2-1} (N, S_T^z) \!+\! \frac{1}{4} N J_2 \!-\! h S_T^z,
\label{eff121}
\end{eqnarray}
where $\varepsilon_{1/2-1} (N, S_T^z)$ denotes the energy per unit cell of the effective mixed spin-($1/2, 1$) Heisenberg chain with unit coupling constant, the number of unit cells $N$ and $z$-component of the total spin $S_T^z$. The mixed spin-($1/2, 1$) Heisenberg chain exhibits at low enough magnetic fields the gapped ferrimagnetic phase with a $z$-component of the total spin $S_T^z = (1 - \frac{1}{2}) \times N = N/2$ due to ordering of energy levels imposed by the Lieb-Mattis theorem \cite{lieb62}, whereby the eigenstates with higher values of the total spin momentum $S_T^z > N/2$ form a continuous band pertinent to the gapless Tomonaga-Luttinger quantum spin liquid emergent at higher magnetic fields when an energy gap above the Lieb-Mattis ferrimagnetic phase closes \cite{silv17,strc17,strk17,silv21,ivan10}. To gain the respective eigenenergies $\varepsilon_{1/2-1} (N, S_T^z)$ of the effective mixed spin-($1/2, 1$) Heisenberg chain one has to resort to state-of-the-art numerical calculations. To this end, we have implemented DMRG simulations of the effective mixed spin-($1/2, 1$) Heisenberg chain with up to 120 spins ($N=60$ unit cells, which correspond to the diamond chain of 180 spins) within the open-source software Algorithms and Libraries for Physics Simulation (ALPS) project \cite{baue11}. Finally, another available ground state relates to the lowest-energy eigenstate of the effective mixed spin-(1/2, 1, 1/2, 0) Heisenberg chain, where singlet and polarized triplet states regularly alternate on vertical dimers. The singlet-dimer states break the effective mixed spin-(1/2, 1, 1/2, 0) Heisenberg chain into noninteracting three-spin clusters, which also turn out to be in the singlet state within the lowest-energy eigenstate with the energy
\begin{eqnarray}
E_{1/2-1-1/2-0} (N, S_T^z = 0) = - \frac{1}{4} N J_2 - N J_1.
\label{eff121120}
\end{eqnarray}
This lowest-energy eigenstate thus corresponds to the singlet tetramer-dimer (TD) phase of the spin-1/2 Heisenberg diamond chain, which is schematically depicted in Fig. \ref{fig:dctd} and mathematically given by the eigenvector 
\begin{widetext}
\begin{eqnarray}
|{\rm TD}\rangle = \prod_{j=1}^{N/2} \frac{1}{\sqrt{3}} \Bigl[&|&\!\!\uparrow_{1,2j-1}\downarrow_{2,2j-1}\uparrow_{3,2j-1}\downarrow_{1,2j}\rangle 
	 + |\!\!\downarrow_{1,2j-1}\uparrow_{2,2j-1}\downarrow_{3,2j-1}\uparrow_{1,2j}\rangle  - \frac{1}{2} \Bigl(|\!\!\uparrow_{1,2j-1}\uparrow_{2,2j-1}\downarrow_{3,2j-1}\downarrow_{1,2j}\rangle \nonumber \\
	&+& |\!\!\uparrow_{1,2j-1}\downarrow_{2,2j-1}\downarrow_{3,2j-1}\uparrow_{1,2j}\rangle 
  + |\!\!\downarrow_{1,2j-1}\uparrow_{2,2j-1}\uparrow_{3,2j-1}\downarrow_{1,2j}\rangle 
	+ |\!\!\downarrow_{1,2j-1}\downarrow_{2,2j-1}\uparrow_{3,2j-1}\uparrow_{1,2j}\rangle \Bigr) \Bigr] \nonumber \\
&\otimes& \frac{1}{\sqrt{2}} \Bigl(|\!\!\uparrow_{2,2j}\downarrow_{3,2j}\rangle - |\!\!\downarrow_{2,2j}\uparrow_{3,2j}\rangle\Bigr).
\label{tdgs}
\end{eqnarray}
\end{widetext}
Note that the singlet TD phase is two-fold degenerate, because another linearly independent eigenstate with the same energy can be obtained from the eigenvector (\ref{tdgs}) by interchanging singlet-tetramer and singlet-dimer states on odd and even unit cells. The TD phase emerges just if the interaction ratio is from the interval $0.91<J_2/J_1<2.0$ and the magnetic field is sufficiently small $h/J_1<0.55$, whereby the highest possible field value for existence of the TH ground state is achieved for $J_2/J_1\approx 1.45$, see Fig.~\ref{fig:gsdc}.

The ground-state phase diagram constructed from all aforedescribed lowest-energy eigenstates of the spin-1/2 Heisenberg diamond chain is depicted in Fig. \ref{fig:gsdc} in the interaction ratio versus magnetic field plane. In total, the displayed phase diagram  involves five different ground states: the gapped Lieb-Mattis ferrimagnetic phase, the gapless Tomonaga-Luttinger quantum spin liquid, the fully polarized ferromagnetic phase, the MD phase and the TD phase. Two horizontal phase boundaries delimiting the Tomonaga-Luttinger quantum spin liquid mark continuous field-driven quantum phase transitions, while all other phase boundaries denote discontinuous field-driven phase transitions accompanied with discontinuous magnetization jumps. The magnetization corresponding to the ferrimagnetic phase and the MD phase equals to one-third of the saturation magnetization, while the TD phase has zero magnetization owing to the singlet character of regularly alternating singlet tetramers and singlet dimers. Contrary to this, the magnetization of the Tomonaga-Luttinger quantum spin liquid continuously rises upon increasing of the magnetic field due to the gapless character of this quantum ground state. To the best of our knowledge, the overall ground-state phase diagram of the symmetric spin-1/2 Heisenberg diamond chain has not been reported in the literature yet, while the zero-field phase boundaries between the ferrimagnetic, TD and MD phases are in a perfect agreement with the former results reported by Takano, Kubo and Sakamoto \cite{taka96}.

\section{Heisenberg diamond chain as monomer-dimer problem}
\label{sec:dcmd}

\begin{figure}
\centering\includegraphics[width=0.9\columnwidth]{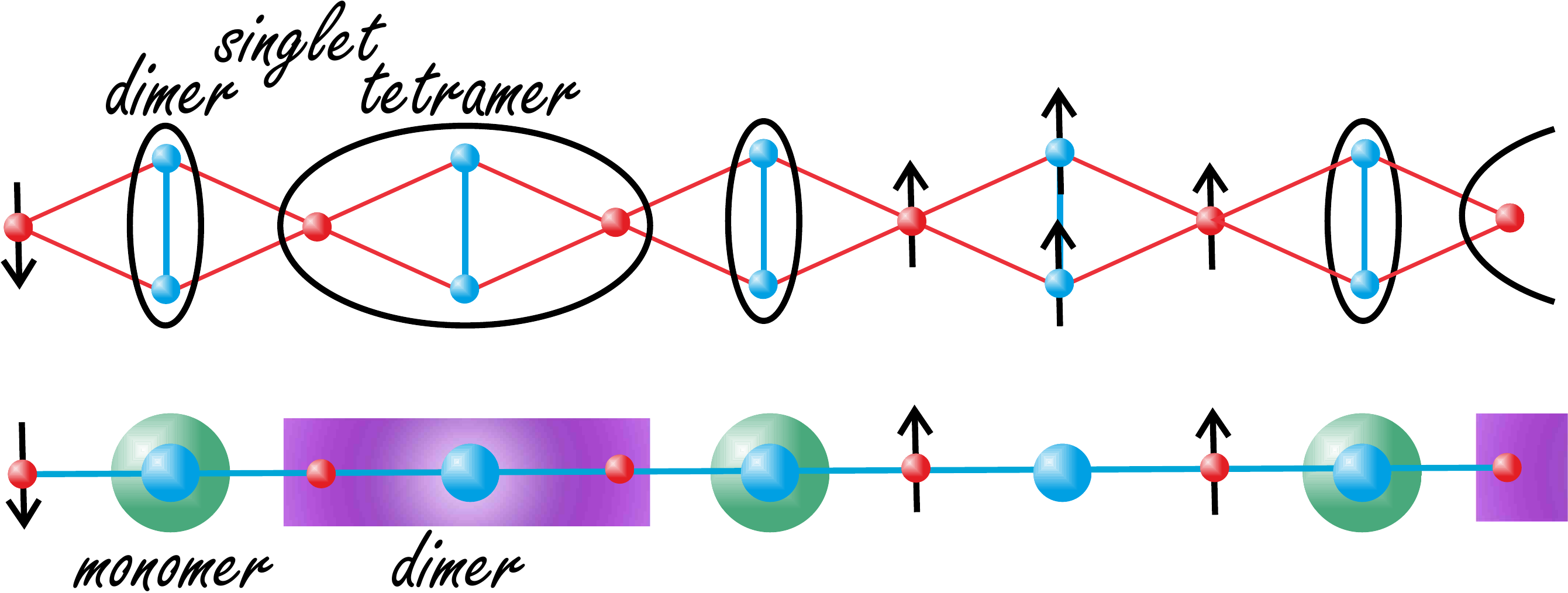}
\vspace{-0.4cm}
\caption{One typical permissible state of the spin-1/2 Heisenberg diamond chain and its equivalent representation within two-component lattice-gas model of hard-core monomers and dimers. Small red spheres of the auxiliary lattice correspond to the monomeric spins $S_{1,j}$, while large blue spheres of the auxiliary lattice correspond to the composite spins $S_{c,j}$ assigned to the vertical dimers $S_{2,j}-S_{3,j}$.}
\label{fig:lmdc}
\end{figure}

In this section, we will turn our attention to a description of low-temperature magnetization curves and thermodynamics of the spin-1/2 Heisenberg diamond chain. It is noteworthy that the standard localized-magnon theory based on a classical lattice-gas model of hard-core monomers offers a satisfactory description of low-temperature magneto-thermodynamics of the spin-1/2 Heisenberg diamond chain just in the highly frustrated parameter region $J_2/J_1>2$ \cite{derz06,derz07,derz12}, whereas our aim is to proceed beyond this simple theory. The low-energy features of the spin-1/2 Heisenberg diamond chain will be effectively described by the one-dimensional lattice-gas model of hard-core monomers and dimers, which will allow us to extend the range of validity of the localized-magnon approach to a less frustrated parameter space involving the singlet TD phase (\ref{tdgs}). The monomer-dimer lattice-gas model defined on an auxiliary one-dimensional lattice includes hard-core monomeric particles assigned to the singlet-dimer states on vertical bonds 
\begin{eqnarray}
|m\rangle_j = \frac{1}{\sqrt{2}} (|\!\!\uparrow_{2,j}\downarrow_{3,j}\rangle - |\!\!\downarrow_{2,j}\uparrow_{3,j}\rangle), 
\label{sd}  
\end{eqnarray}
as well as, hard-core dimeric particles assigned to the singlet-tetramer states of diamond (four-spin) clusters being a cornerstone of the TD ground state (\ref{tdgs}) 
\begin{eqnarray}
|d\rangle_j = 
  \frac{1}{\sqrt{3}}\Bigl(|\!\!\uparrow_{1,j}\downarrow_{2,j}\uparrow_{3,j}\downarrow_{1,j+1}\rangle &+& |\!\!\downarrow_{1,j}\uparrow_{2,j}\downarrow_{3,j}\uparrow_{1,j+1}\rangle\Bigr)   \nonumber \\
- \frac{1}{\sqrt{12}} \Bigl(|\!\!\uparrow_{1,j}\uparrow_{2,j}\downarrow_{3,j}\downarrow_{1,j+1}\rangle &+& |\!\!\uparrow_{1,j}\downarrow_{2,j}\downarrow_{3,j}\uparrow_{1,j+1}\rangle \nonumber \\
+ |\!\!\downarrow_{1,j}\uparrow_{2,j}\uparrow_{3,j}\downarrow_{1,j+1}\rangle  &+& |\!\!\downarrow_{1,j}\downarrow_{2,j}\uparrow_{3,j}\uparrow_{1,j+1}\rangle\Bigr). \nonumber \\
\label{st}
\end{eqnarray}
One typical permissible state of the spin-1/2 Heisenberg diamond chain and its equivalent representation within the developed monomer-dimer lattice-gas model is displayed in Fig. \ref{fig:lmdc}. Small red circles of the auxiliary lattice correspond to lattice sites of the monomeric spins $S_{1,j}$ and large blue circles of the auxiliary lattice correspond to lattice positions of the composite spins $S_{c,j} = S_{2,j}+S_{3,j}$ of the vertical dimers. The composite spins corresponding to the vertical dimers are accessible to the fully polarized ferromagnetic state acting as a reference vacuum state, the singlet-dimer state (\ref{sd}) represented by a monomeric particle schematically shown by a large green circle and the singlet-tetramer state (\ref{st}) represented by a dimeric particle (large violet rectangle) incorporating two enclosing monomeric spins $S_{1,j}$ and $S_{1,j+1}$ as well. If the energy is defined relative with respect to the fully polarized ferromagnetic state serving as a reference state, then, one should assign the chemical potential $\mu_1 = J_1 + J_2 - h$ to the monomeric particles connected with the singlet-dimer state (\ref{sd}) and the chemical potential $\mu_2 = 3J_1 - h$ to the dimeric particles associated with the singlet-tetramer state (\ref{st}). 
The effective Hamiltonian of the  one-dimensional lattice-gas model of hard-core monomers and dimers consequently reads
\begin{eqnarray}
{\cal H}_{\rm eff} = E_{\rm FM}^0 &-& h \sum_{j=1}^N [1 + S_{1,j}^z (1-d_{j-1})(1-d_{j})] \nonumber \\
&-& \mu_1 \sum_{j=1}^{N} m_{j} - \mu_2 \sum_{j=1}^{N} d_{j},
\label{lgdm}
\end{eqnarray}
where $E_{\rm FM}^0 = NJ_1 + N J_2/4$ refers to the zero-field energy of the fully polarized ferromagnetic state, the second term is the respective Zeeman's energy; 
$m_j= 0,1$ and $d_j=0,1$ are occupation numbers  of the hard-core monomeric and dimeric particles, respectively. The partition function of the monomer-dimer lattice-gas model defined through the Hamiltonian (\ref{lgdm}) can be written in the compact form
\begin{eqnarray}
{\cal Z} &=& {\rm e}^{-\beta E_{\rm FM}^0} \!\!
\sum_{\{S_{1,j}^z\}} \! \sum_{\{m_{j}\}} \! \sum_{\{d_{j}\}} \! 
\prod_{j=1}^N  \! \frac{1}{4^{d_j}} (1{-}d_{j-1} d_{j}) (1 {-} m_j d_j)   \nonumber \\
&\times& \exp \{ \beta [\mu_1 m_{j} {+} \mu_2 d_{j} {+} h S_{1,j}^z (1{-}d_{j-1})(1{-}d_{j})] \}.
\label{dcpf}
\end{eqnarray}
Here, $\beta=1/(k_{\rm_{B}}T)$, $k_{\rm{B}}$ is the Boltzmann constant, $T$ is absolute temperature, the summation $\sum_{\{S_{1,j}^z\}}$ runs over spin states of all monomeric spins $S_{1,j}$, the summations $\sum_{\{m_{j}\}}$ and $\sum_{\{d_{j}\}}$ are carried out over all possible values of the respective occupation numbers. The factor $(1-d_{j-1} d_{j}) (1 - m_j d_j)$ ensures a hard-core constraint forbidding overlap and/or double occupancy of auxiliary lattice sites by the monomeric and dimeric particles, while the factor $\frac{1}{4^{d_j}}$ is the correction term for the singlet-tetramer state (\ref{st}) that would be without this factor accounted four times when one performs the summation over spin states of two monomeric spins $S_{1,j}$ and $S_{1,j+1}$. The important feature of the partition function (\ref{dcpf}) is the possibility to sum over all the states of the monomeric spins $S_{1,j}$ and hard-core monomer particles $m_j$ independently. In fact, the complete hard-core conditions between the quasi-particles should also include the projection operator $[1-(1-m_i)d_{i+1}][1-d_{i}(1-m_{i+1})]$, which ensures that the singlet-plaquette state ($d_i$ particle) cannot be followed by the fully polarized state on the adjacent dimer. The energy of the latter state is much larger that the energy scale of allowed states present in the effective Hamiltonian (\ref{lgdm}) and for simplicity, it can be completely left out from consideration. After performing the summations $\sum_{\{S_{1,j}^z\}}$ and  $\sum_{\{m_{j}\}}$ the expression behind the product symbol can be expressed solely in terms of the occupation numbers of the dimeric particles  
\begin{eqnarray}
\mbox{T} (d_{j}, d_{j+1}) \!&=&\! \frac{1}{4^{d_j}} (1{-}d_{j-1}d_{j}) 2 {\rm e}^{\beta h + \beta \mu_2 d_j} [1 {+} (1{-}d_j){\rm e}^{\beta \mu_1}] \nonumber \\ 
&\times& \cosh \!\left[\!\frac{\beta h}{2} (1{-}d_{j-1})(1{-}d_j)\!\right].
\label{dmtm}
\end{eqnarray}
The expression (\ref{dmtm}) can be in turn identified as the transfer matrix, which allows a simple calculation of the partition function (\ref{dcpf}) within the transfer-matrix method \cite{kram41} exploiting a consecutive summation over the occupation numbers of the dimeric particles 
\begin{eqnarray}
{\cal Z} &=& {\rm e}^{-\beta E_{\rm FM}^0} \sum_{\{d_{j}\}} 
\prod_{j=1}^N  \mbox{T} (d_{j-1}, d_{j}) 
                    = {\rm e}^{-\beta E_{\rm FM}^0} \, \mbox{Tr} \, \mbox{T}^N  \nonumber \\
										&=& {\rm e}^{-\beta E_{\rm FM}^0} (\lambda_{+}^N + \lambda_{-}^N). 
\label{pfdc}
\end{eqnarray}
The final expression for the partition function (\ref{pfdc}) is given through two eigenvalues $\lambda_{\pm}$ obtained after a straightforward diagonalization of two-by-two transfer matrix (\ref{dmtm})
\begin{eqnarray}
\lambda_{\pm} &=& {\rm e}^{\beta h}\Biggl\{ \cosh\!\left(\!\frac{\beta h}{2}\!\right)\! 
\Xi_d \pm \sqrt{\left[\cosh\!\left(\!\frac{\beta h}{2}\!\right)\! 
\Xi_d \right]^2 \!\!\! + {\rm e}^{\beta \mu_2} \Xi_d} \Biggr\}, \nonumber \\ \Xi_d &=& 1 + {\rm e}^{\beta \mu_1}.
\label{chid}
\end{eqnarray}
The free-energy density of the finite-size spin-1/2 Heisenberg diamond chain normalized per spin consequently reads
\begin{eqnarray}
f_{3N} &=& - k_{\rm B} T \frac{1}{3N} \ln {\cal Z} \nonumber \\ 
&=& \frac{1}{3} \left(J_1 + \frac{J_2}{4} \right) - \frac{1}{3N} k_{\rm B} T \ln (\lambda_{+}^N + \lambda_{-}^N).
\label{fed}
\end{eqnarray}
The simpler expression can be acquired for the free-energy density of the infinite spin-1/2 Heisenberg diamond chain, which depends in the thermodynamic limit $N \to \infty$ solely on the larger  eigenvalue of the transfer-matrix
\begin{eqnarray}
f_{\infty} = - k_{\rm B} T \lim_{N \to \infty} \frac{1}{3N} \ln {\cal Z} = \frac{1}{3} \left(J_1 {+} \frac{J_2}{4} \right) {-} \frac{1}{3} k_{\rm B} T \ln \lambda_{+}. \nonumber \\
\label{fedn}
\end{eqnarray}
It is noteworthy that the final formulas (\ref{fed}) and (\ref{fedn}) for the free-energy density allow a straightforward calculation of the magnetization, susceptibility, entropy and specific heat for a finite and infinite spin-1/2 Heisenberg diamond chain, respectively. To verify a reliability of the effective description based on the monomer-dimer lattice-gas model (\ref{lgdm}) we will compare as-obtained magnetization and specific-heat data with the extensive numerical calculations employing the ED and FTLM implemented within the open-source ALPS \cite{baue11} and Spinpack \cite{rich10,schu10} softwares. 

\begin{figure*}
   \includegraphics[width=0.49\textwidth]{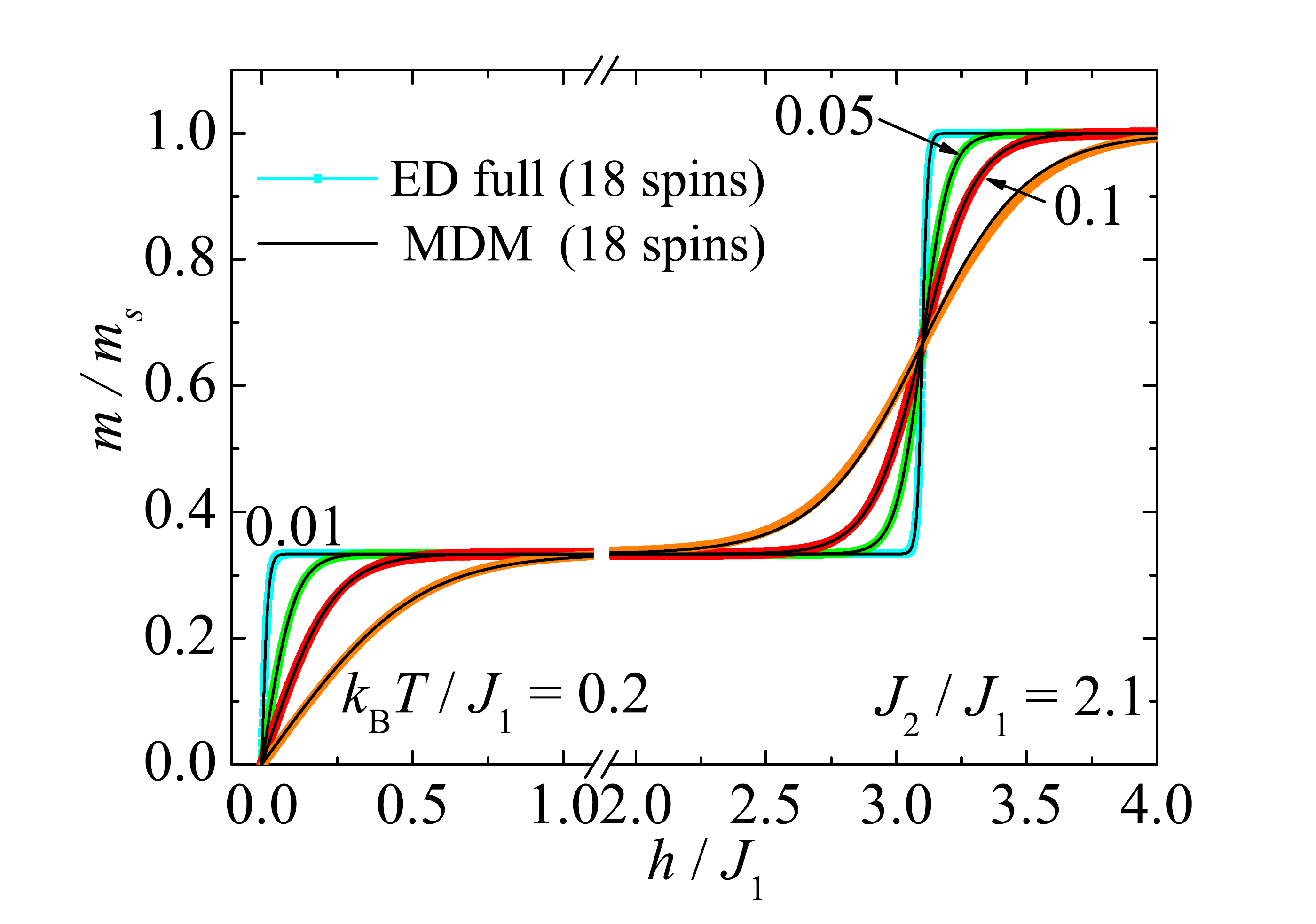}
  \hspace*{0.0cm}
  \includegraphics[width=0.49\textwidth]{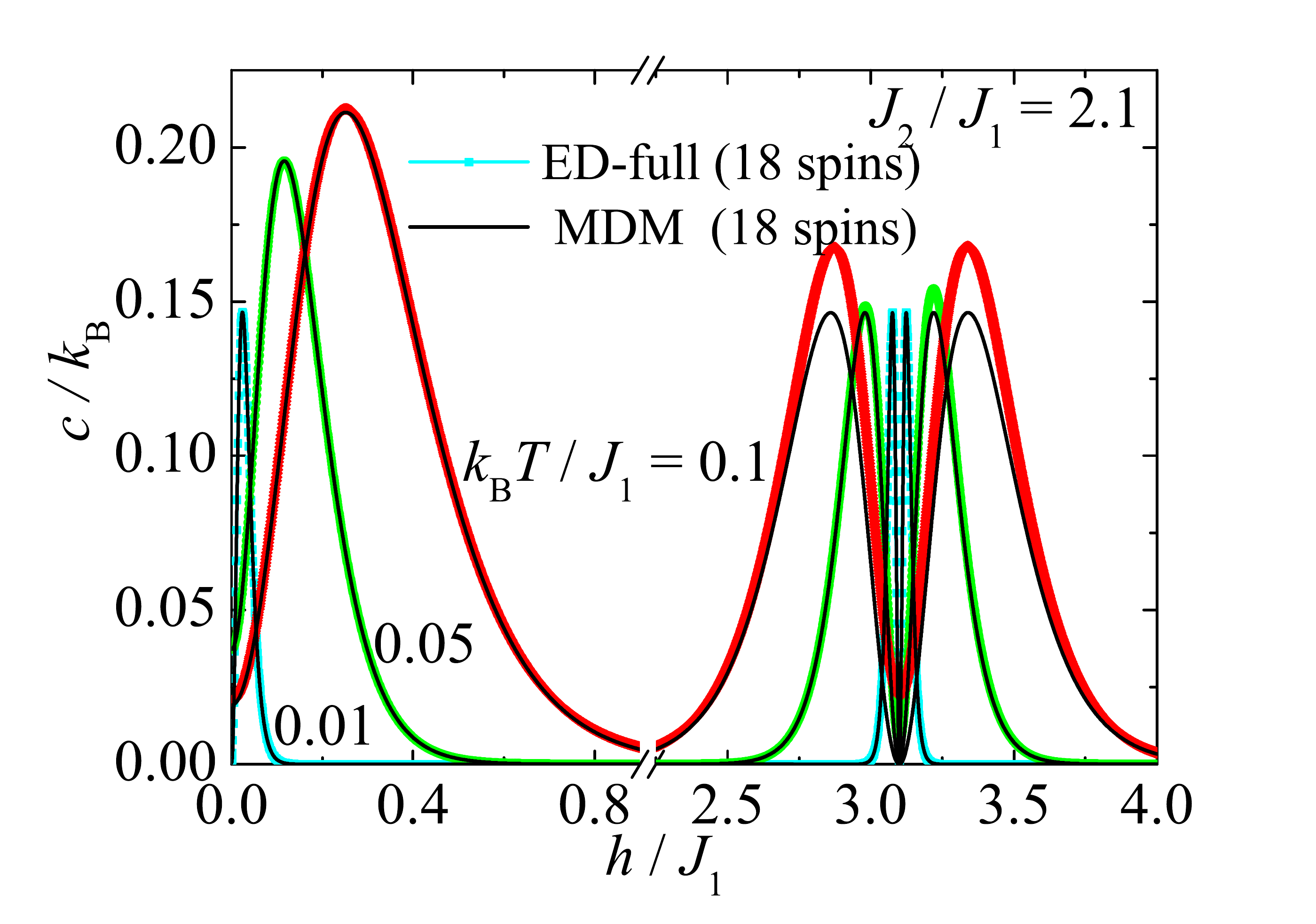} \\  
	\vspace*{-0.2cm}
  \includegraphics[width=0.49\textwidth]{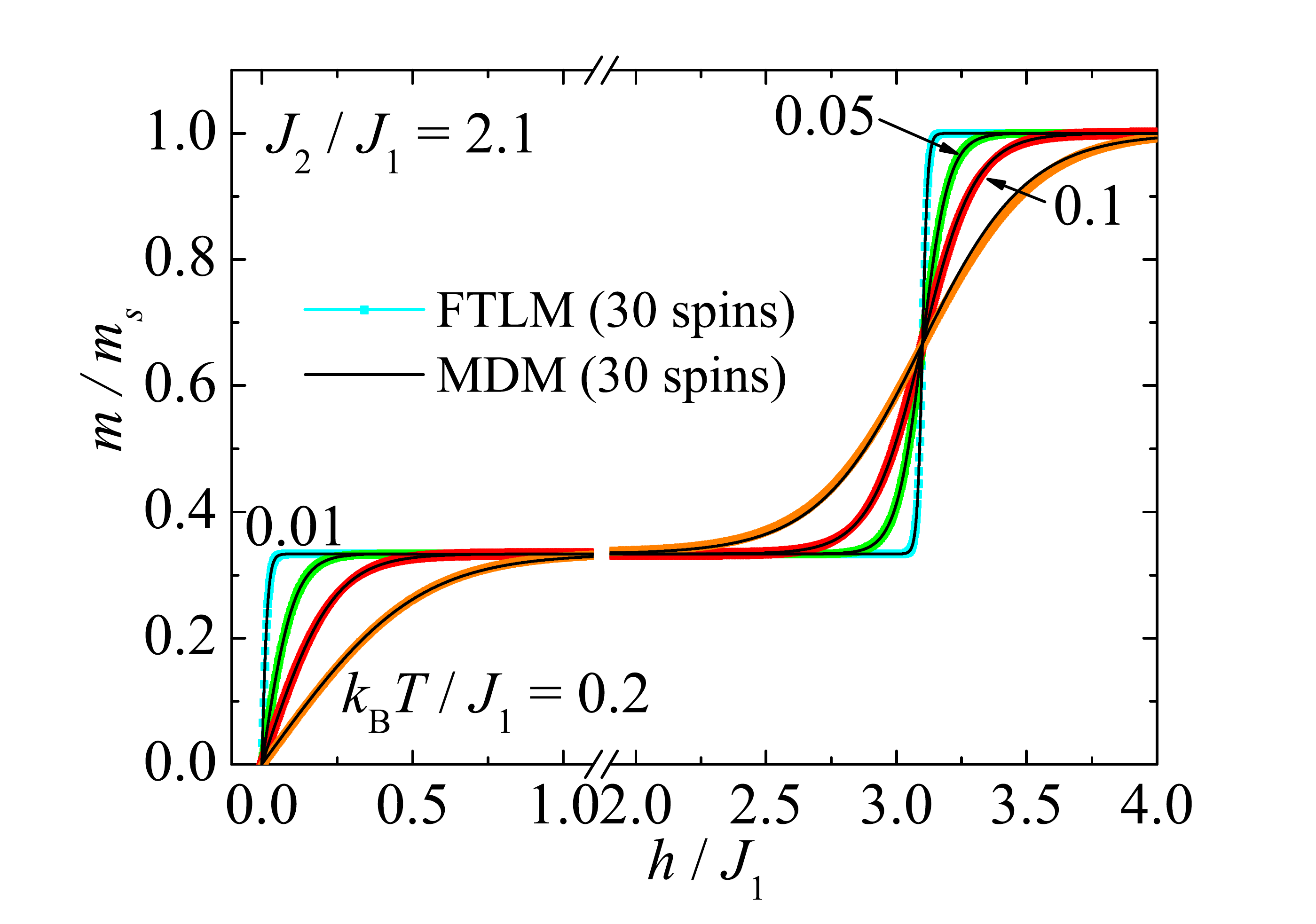}
  \hspace*{0.0cm}
  \includegraphics[width=0.49\textwidth]{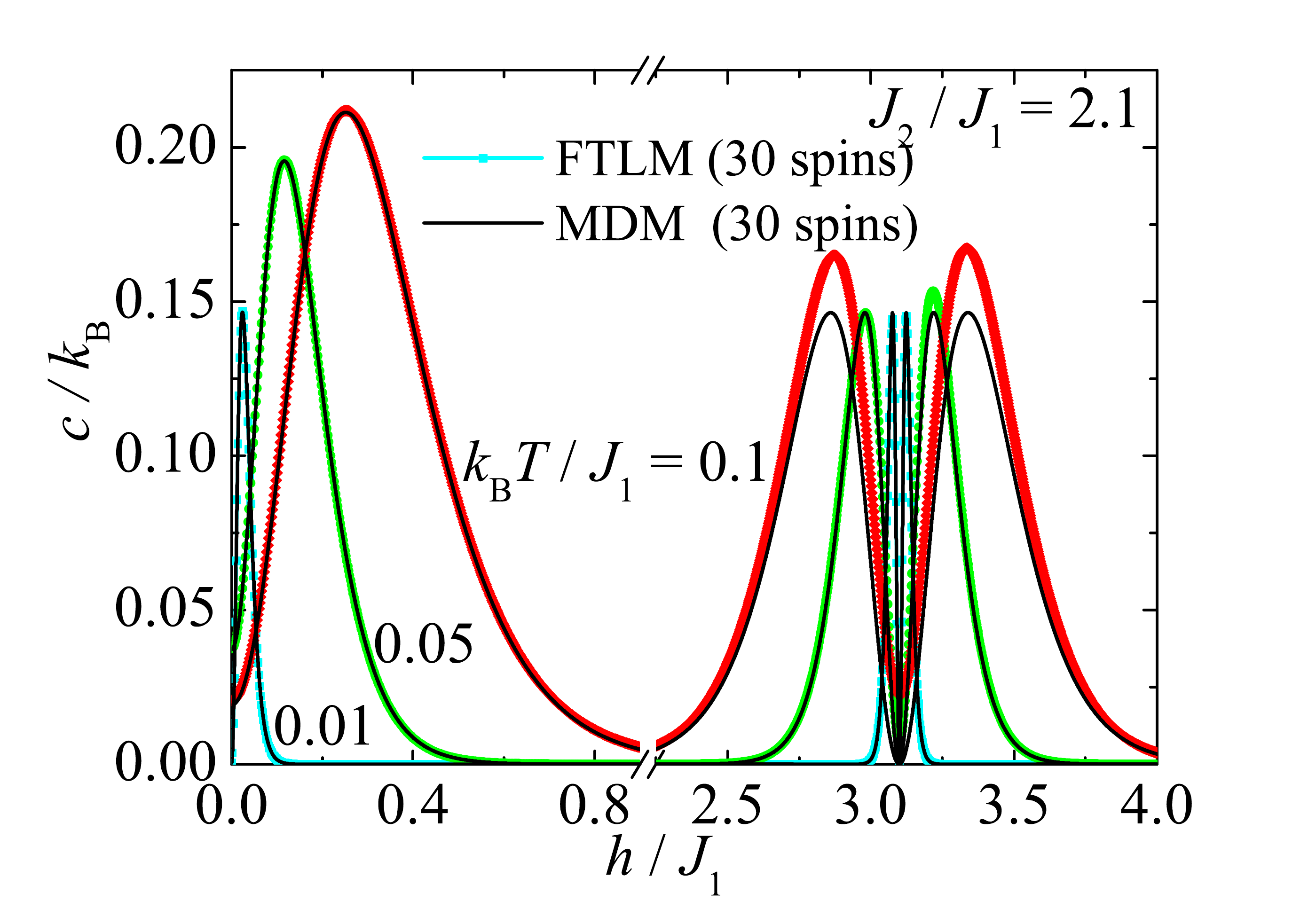}
	\vspace*{-0.4cm}
	\caption{(Left panel) Magnetization curves of the spin-1/2 Heisenberg diamond chain with $J_2/J_1=2.1$ obtained from the effective monomer-dimer model (MDM) versus the ED data for 18 spins (upper panel) and the FTLM data for 30 spins (lower panel) of the original model (\ref{hamdc}). (Right panel) The same as in the left panel but for the specific heat. Thin black lines show analytical results derived from the effective monomer-dimer model (\ref{lgdm}), while colored symbols refer to the ED (upper panel) or FTLM (lower panel). Note that there is an axis break in the middle of intermediate 1/3-plateau.}
\label{prvasadadc}
\end{figure*}	
	
First, let us comment the most interesting results for the highly frustrated spin-1/2 Heisenberg diamond chain with $J_2/J_1 > 2$. For illustrative purposes, magnetization curves of the spin-1/2 Heisenberg diamond chain are displayed in the left panel of Fig. \ref{prvasadadc} for the fixed value of the interaction ratio $J_2/J_1=2.1$ and four different temperatures. The magnetization data derived from the finite-size formula (\ref{fed}) for the free-energy density of the effective monomer-dimer model (\ref{lgdm}) are compared in this figure with the full ED data for 18 spins ($N=6$ unit cells, upper panel) and FTLM results for 30 spins ($N=10$ unit cells, lower panel) of the spin-1/2 Heisenberg diamond chain (\ref{hamdc}). As one can see, the magnetization curves acquired from the monomer-dimer model (\ref{lgdm}) perfectly coincide at sufficiently low temperatures $k_{\rm B} T/J_1 \lesssim 0.2$ with accurate numerical results for the magnetization curves of the spin-1/2 Heisenberg diamond chain. Generally, the intermediate 1/3-plateau resembling the MD ground state gradually shrinks upon increasing of temperature and the magnetization curve becomes smoother. To gain a more complete understanding, the magnetic-field variations of the specific heat of the spin-1/2 Heisenberg diamond chain are depicted in the right panel of Fig. \ref{prvasadadc} for the same fixed value of the interaction ratio $J_2/J_1=2.1$ and three different temperatures. It is obvious from this figure that the perfect agreement between the specific-heat data stemming from the effective monomer-dimer model (\ref{lgdm}) and the precise numerical data acquired for the spin-1/2 Heisenberg diamond chain (\ref{hamdc}) is limited to much lower temperatures $k_{\rm{B}}T/J_1\lesssim 0.05$, while the discrepancy becomes more pronounced at higher temperatures even if the effective description still at least qualitatively reproduces the most essential features of the accurate numerical data. It should be also pointed out that the magnetization and specific heat of the spin-1/2 Heisenberg diamond chain do not show at low enough temperatures almost any finite-size dependence (confront data for the diamond chain with 18 and 30 spins), because most thermally populated excited states are from the monomeric universality class. 

\begin{figure*}
  \includegraphics[width=0.49\textwidth]{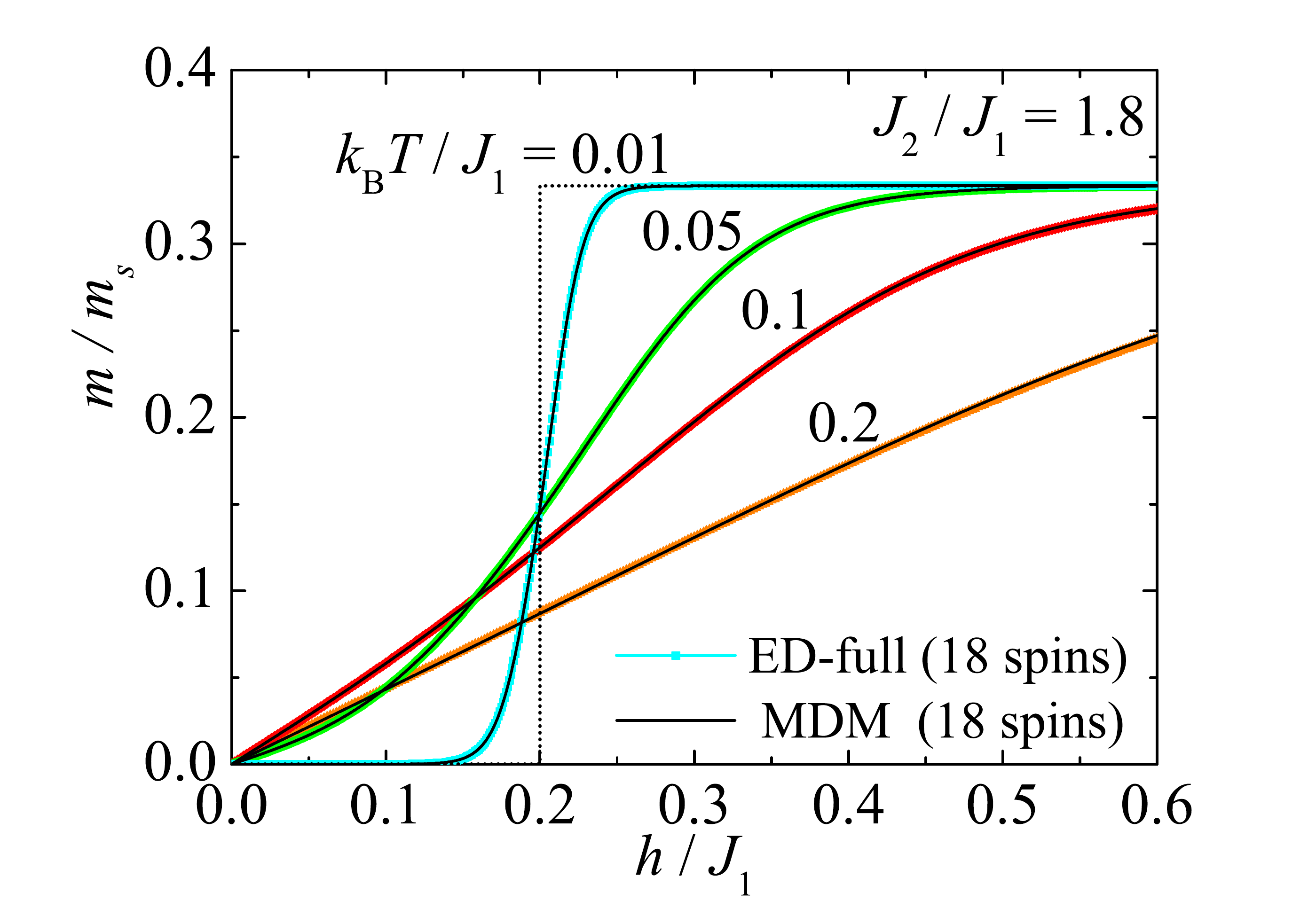}
  \hspace*{0.0cm}
  \includegraphics[width=0.49\textwidth]{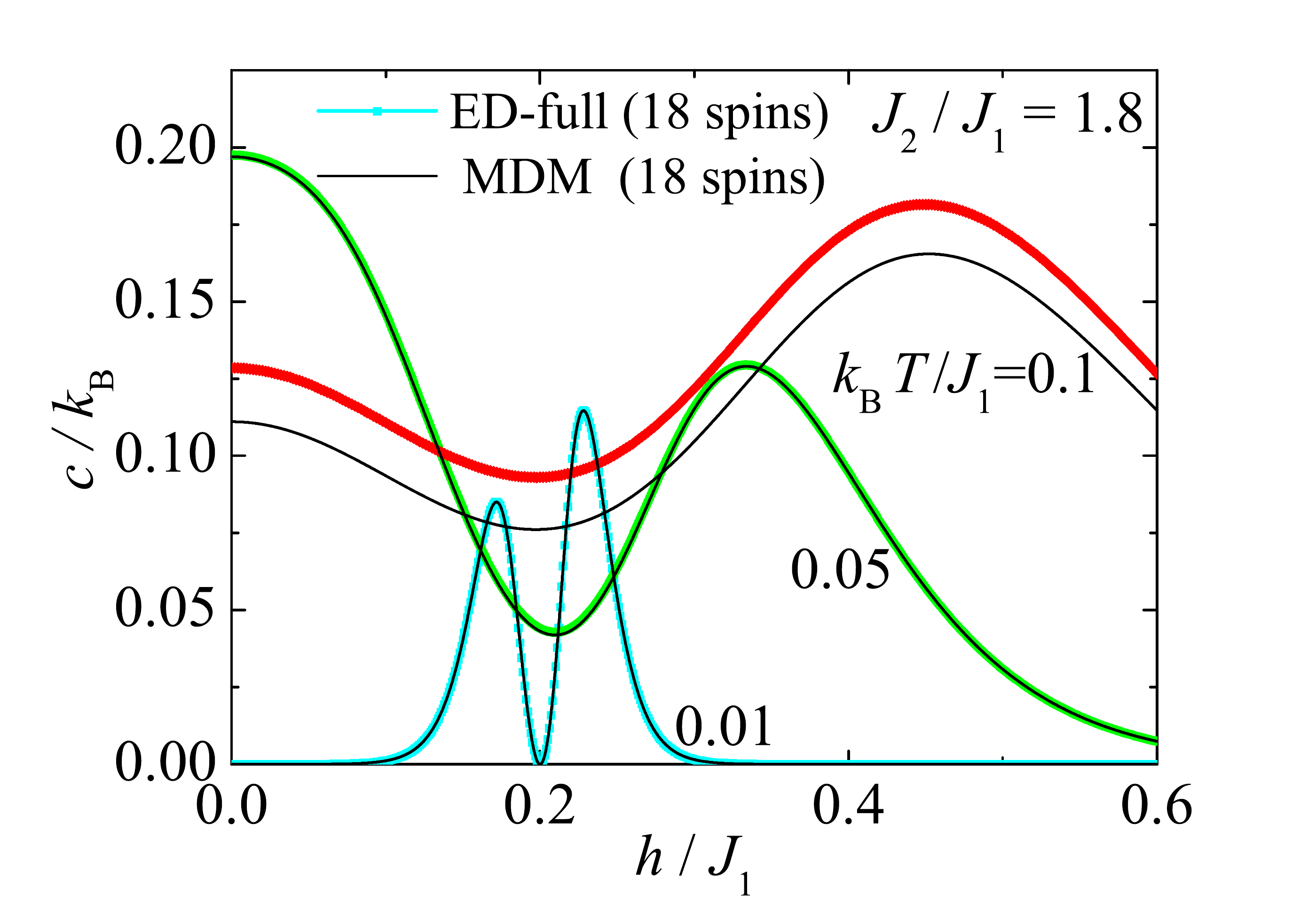}  
	\vspace*{-0.2cm}
  \includegraphics[width=0.49\textwidth]{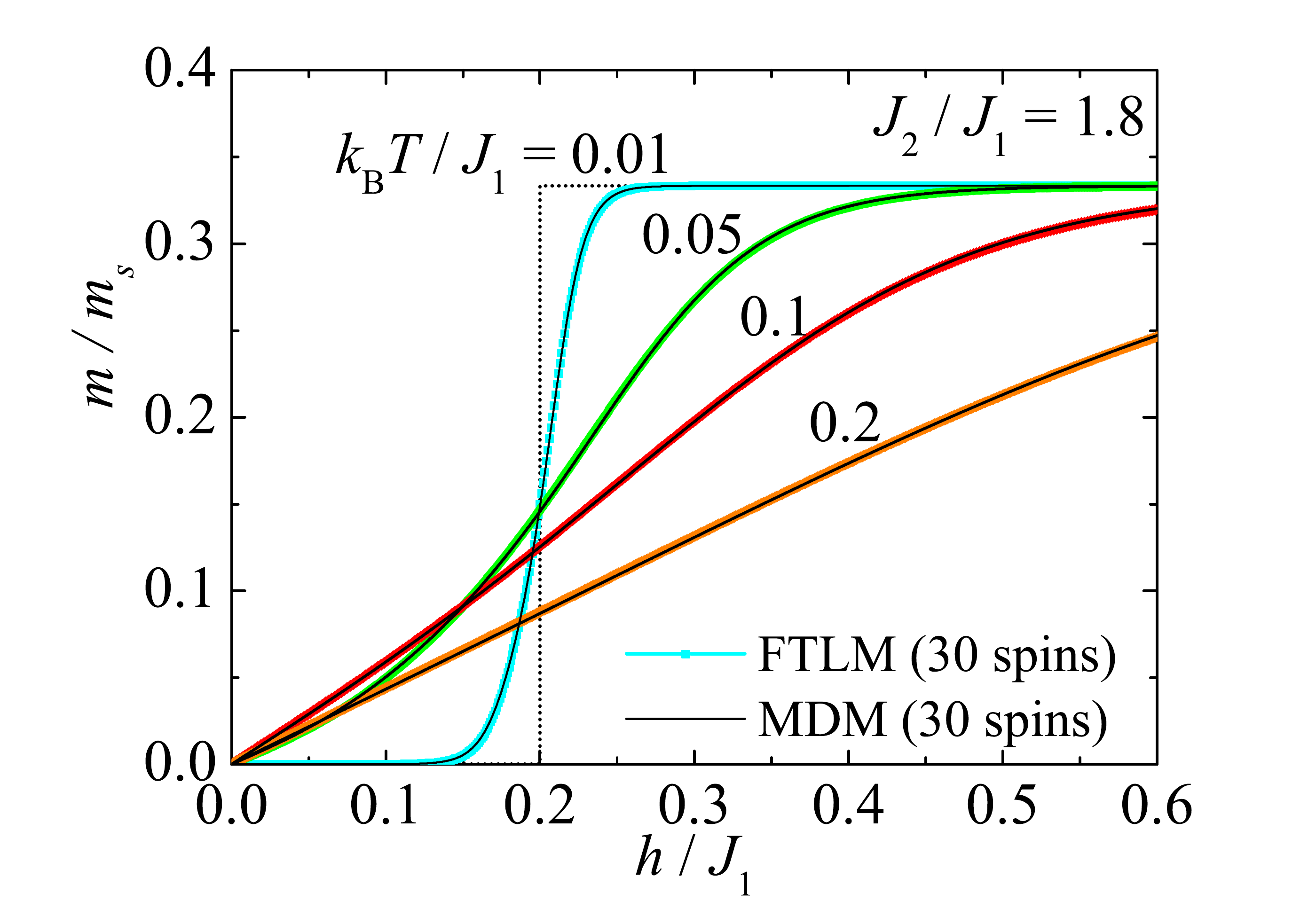}
  \hspace*{0.0cm}
  \includegraphics[width=0.49\textwidth]{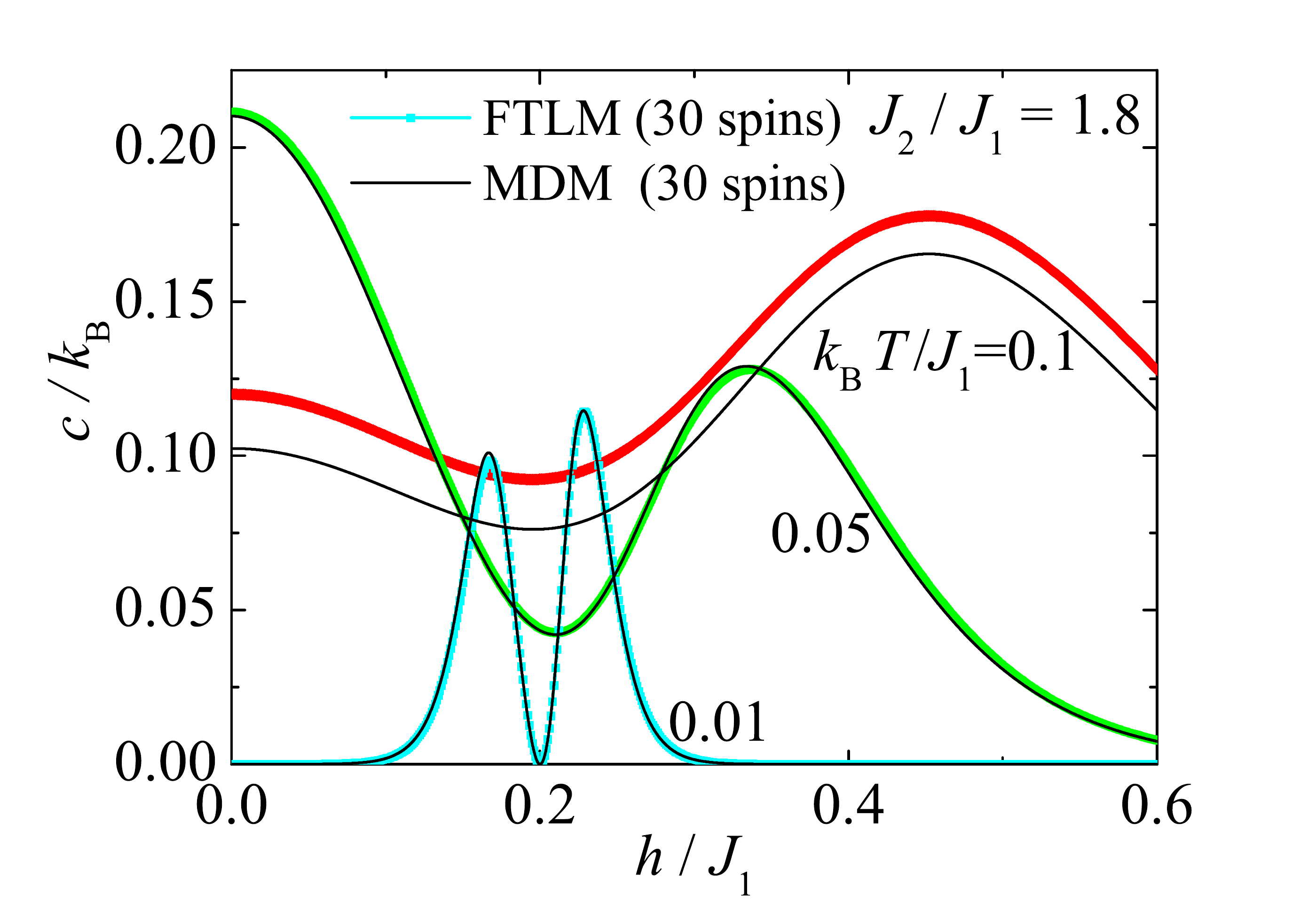}
	\vspace*{-0.4cm}
	\caption{(Left panel) Magnetization curves of the spin-1/2 Heisenberg diamond chain with $J_2/J_1=1.8$ obtained from the effective monomer-dimer model (MDM) versus the ED data for 18 spins (upper panel) and the FTLM data for 30 spins (lower panel) of the original model (\ref{hamdc}). (Right panel) The same as in the left panel but for the specific heat. Thin black lines show analytical results derived from the effective monomer-dimer model (\ref{lgdm}), while colored symbols refer to the ED (upper panel) or FTLM (lower panel).}
	\label{fig:dcless}
\end{figure*}
Magnetization process and specific heat are displayed in Fig. \ref{fig:dcless} for the spin-1/2 Heiseberg diamond chain with a relative strength of the interaction constants $J_2/J_1=1.8$ promoting a moderately strong spin frustration. Under this condition, the spin-1/2 Heisenberg diamond chain exhibits in a low-field region zero and one-third magnetization plateaus, which can be ascribed to the TD and MD ground states, respectively. The magnetization curves extracted from the effective lattice-gas model of hard-core monomers and dimers (\ref{lgdm}) are in excellent agreement with the precise numerical data obtained from the full ED (upper panel) or FTLM (lower panel) up to moderate temperatures $k_{\rm{B}}T/J_1\lesssim 0.2$, see the left panel in Fig. \ref{fig:dcless}. The magnetic-field dependences of the specific heat of the spin-1/2 Heisenberg diamond chain are plotted on the right panel  of Fig. \ref{fig:dcless} for the same value of the interaction ratio $J_2/J_1=1.8$ and three different temperatures. It is evident that the specific heat displays in a low-field region a pronounced double-peak dependence, which appears in a vicinity of the field-driven phase transition between the TD and MD ground states emergent at $h/J_1 = 0.2$. Unlike the previous case, the peak heights of the specific heat above and below the relevant field-induced phase transition are not the same due to difference in the relative degeneracy of the TD and MD ground states. It turns out that the low-temperature dependences of the specific heat stemming from the effective monomer-dimer model (\ref{lgdm}) satisfactorily reproduce the accurate numerical data gained for the spin-1/2 Heisenberg diamond chain (\ref{hamdc}) with the help of ED and FTLM only at low enough temperatures $k_{\rm{B}}T/J_1 \lesssim 0.05$, whereas there is only qualitative rather than quantitative agreement at higher temperatures $k_{\rm{B}}T/J_1 \gtrsim 0.1$. Another interesting observation is that the less frustrated spin-1/2 Heisenberg diamond chain with $J_2/J_1=1.8$ shows a marked finite-size effect of the specific heat when both peak heights are at sufficiently low temperatures somewhat higher for the diamond chain with 30 spins than for 18 spins. 		
		
\section{Ground states of a spin-1/2 Heisenberg octahedral chain}
\label{sec:oc}
\begin{figure}
\centering\includegraphics[width=0.9\columnwidth]{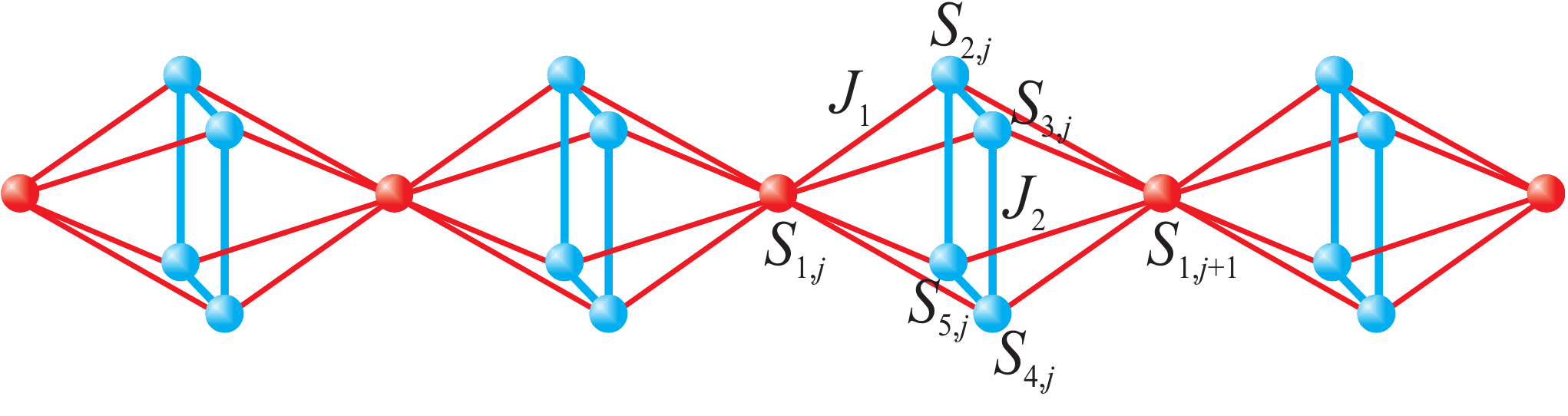}
\vspace{-0.4cm}
 \caption{A schematic illustration of the spin-1/2 Heisenberg octahedral chain including
notation for the lattice sites and assumed coupling constants.}
\label{fig:oc}
\end{figure}
Another paradigmatic example to be considered hereafter is the spin-1/2 Heisenberg octahedral chain schematically shown in Fig. \ref{fig:oc} and defined via the Hamiltonian
\begin{eqnarray}
\hat{\cal H} &=& 
\sum_{j=1}^{N} \Bigl[ J_1 (\boldsymbol{\hat{S}}_{1,j} + \boldsymbol{\hat{S}}_{1,j+1}) \!\cdot\! (\boldsymbol{\hat{S}}_{2,j} + \boldsymbol{\hat{S}}_{3,j} + \boldsymbol{\hat{S}}_{4,j} + \boldsymbol{\hat{S}}_{5,j}) \Bigr.  \nonumber \\
&+& J_2 (\boldsymbol{\hat{S}}_{2,j}\!\cdot\!\boldsymbol{\hat{S}}_{3,j} + \boldsymbol{\hat{S}}_{3,j}\!\cdot\!\boldsymbol{\hat{S}}_{4,j}
+ \boldsymbol{\hat{S}}_{4,j}\!\cdot\!\boldsymbol{\hat{S}}_{5,j} + \boldsymbol{\hat{S}}_{5,j}\!\cdot\!\boldsymbol{\hat{S}}_{2,j}) \nonumber \\
&-& h \sum_{i=1}^{5} \hat{S}_{i,j}^{z} \Bigr].
\label{hamoc}
\end{eqnarray}
Here, the coupling constant $J_2$ denotes antiferromagnetic exchange interaction between nearest-neighbor spins from square plaquettes and the coupling constant $J_1$ accounts for the nearest-neighbor interaction between monomeric spins and square-plaquette spins. The parameter $h$ stands for the external magnetic field, $N$ is the total number of unit cells and the periodic boundary condition is assumed $S_{1,N}\equiv S_{1,1}$ for simplicity. A full ground-state phase diagram of the spin-1/2 Heisenberg octahedral chain (\ref{hamoc}) was established using the variational technique, localized-magnon approach and DMRG calculations in our previous paper \cite{stre17}, to which readers interested in further calculation details are referred to. To make our subsequent discussion self-contained  we will merely quote only the main outcomes of this calculation procedure, which was accomplished in a similar fashion as thoroughly described for the spin-1/2 Heisenberg diamond chain in the preceding part.

\begin{figure}
\centering\includegraphics[width=0.9\columnwidth]{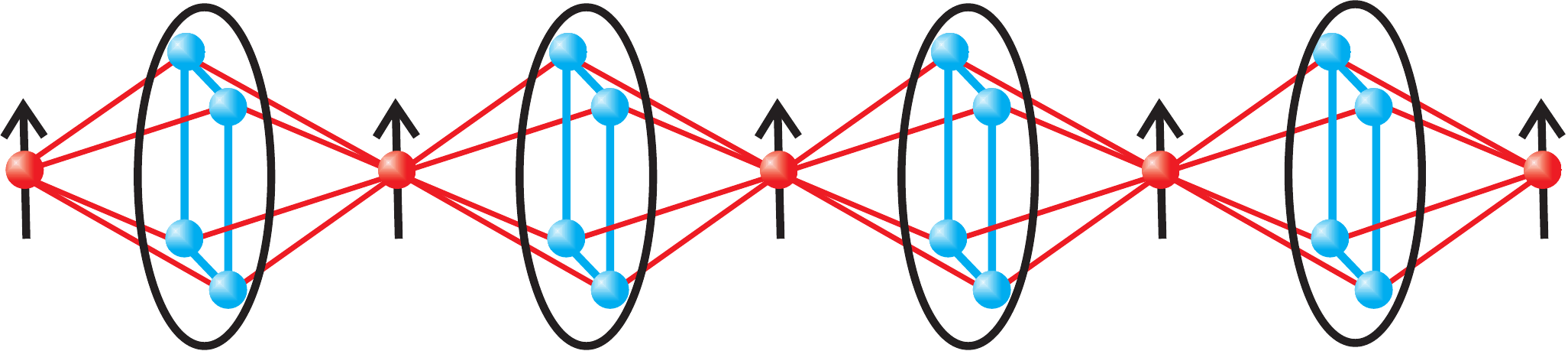} 
\vspace{-0.4cm}
\caption{A schematic illustration of the monomer-tetramer ground state (\ref{vecMT}), which is responsible for occurrence of the intermediate one-fifth plateau in a zero-temperature magnetization curve. An oval denotes a singlet-tetramer state.}
\label{fig:ocmt}
\end{figure}

The variational method provides for the spin-1/2 Heisenberg octahedral chain (\ref{hamoc}) an exact evidence of the monomer-tetramer (MT) ground state in the highly frustrated parameter region $J_2>2J_1$ and low enough magnetic fields $h<J_1+J_2$. The MT ground state is for better illustration schematically drawn in Fig. \ref{fig:ocmt} and can be defined through the eigenvector with character of a tensor-product state of uncorrelated monomeric spins and singlet tetramers on square plaquettes
\begin{widetext}
\begin{eqnarray}
|{\rm MT} \rangle = \left\{ \begin{array}{l} \displaystyle \prod_{j=1}^N \! |{S}_{1,j}\rangle \!\otimes\! 
\frac{1}{\sqrt{3}} \Bigl[\frac{1}{2} \Bigl( |\!\uparrow_{2,j}\uparrow_{3,j}\downarrow_{4,j}\downarrow_{5,j}\rangle + |\!\uparrow_{2,j}\downarrow_{3,j}\downarrow_{4,j}\uparrow_{5,j}\rangle 
+ |\!\downarrow_{2,j}\uparrow_{3,j}\uparrow_{4,j}\downarrow_{5,j}\rangle + |\!\downarrow_{2,j}\downarrow_{3,j}\uparrow_{4,j}\uparrow_{5,j}\rangle \Bigr) \\ 
\hfill
- |\!\uparrow_{2,j}\downarrow_{3,j}\uparrow_{4,j}\downarrow_{5,j}\rangle - |\!\downarrow_{2,j}\uparrow_{3,j}\downarrow_{4,j}\uparrow_{5,j}\rangle  \Bigr], \quad h = 0, \\
\displaystyle \prod_{j=1}^N \! |\!\uparrow_{1,j}\rangle \!\otimes\! 
\frac{1}{\sqrt{3}} \Bigl[\frac{1}{2} \Bigl(|\!\uparrow_{2,j}\uparrow_{3,j}\downarrow_{4,j}\downarrow_{5,j}\rangle + |\!\uparrow_{2,j}\downarrow_{3,j}\downarrow_{4,j}\uparrow_{5,j}\rangle 
+ |\!\downarrow_{2,j}\uparrow_{3,j}\uparrow_{4,j}\downarrow_{5,j}\rangle + |\!\downarrow_{2,j}\downarrow_{3,j}\uparrow_{4,j}\uparrow_{5,j}\rangle \Bigr)  \quad \qquad \qquad \qquad \\
\hfill
- |\!\uparrow_{2,j}\downarrow_{3,j}\uparrow_{4,j}\downarrow_{5,j}\rangle - |\!\downarrow_{2,j}\uparrow_{3,j}\downarrow_{4,j}\uparrow_{5,j}\rangle \Bigr], \quad h > 0,
\end{array} \right.
\label{vecMT}
\end{eqnarray}
\end{widetext}
where $|{S}_{1,j}\rangle$ denotes any out of two available states $|\!\!\uparrow_{1,j}\rangle$ and $|\!\!\downarrow_{1,j}\rangle$ of the monomeric spins. It is noteworthy that the monomeric spins are within the MT phase (\ref{vecMT}) completely free to flip at zero magnetic field due to a spin frustration invoked by a singlet-tetramer state, which can be alternatively viewed as a bound two-magnon eigenstate of a square plaquette. Owing to this fact, the monomeric spins become fully polarized by any nonzero external magnetic field and the MT ground state (\ref{vecMT}) will consequently manifest itself in a zero-temperature magnetization curve as the intermediate one-fifth plateau. 

\begin{figure}
\centering\includegraphics[width=0.9\columnwidth]{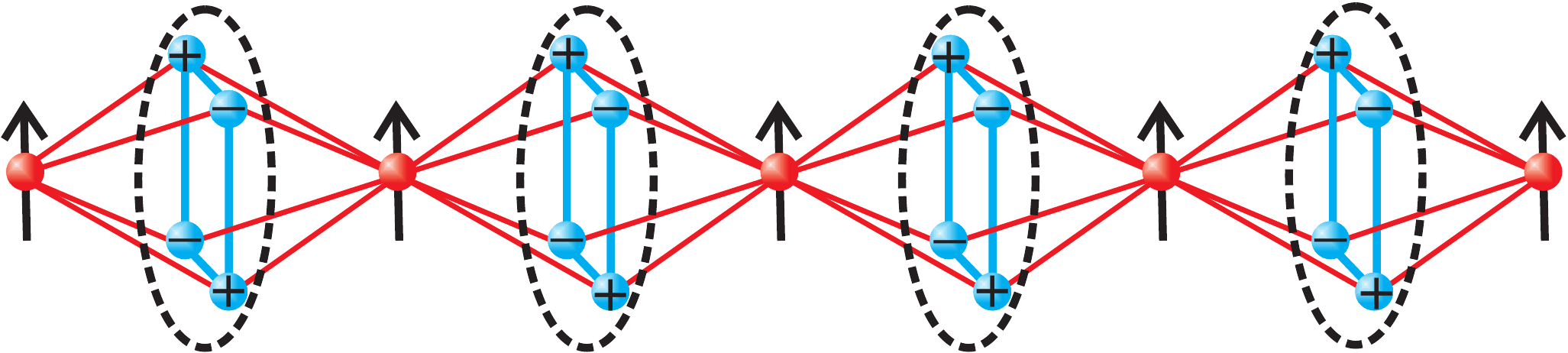} 
\vspace{-0.4cm}
\caption{A schematic illustration of the magnon-crystal ground state (\ref{vecMC}), which is responsible for occurrence of the intermediate three-fifths plateau in a zero-temperature magnetization curve. An oval denotes a magnon bound to a square plaquette due to a destructive quantum interference caused by alternating signs of the probability amplitudes.}
\label{fig:ocbm}
\end{figure}

On the other hand, the localized-magnon theory developed for the spin-1/2 Heisenberg octahedral chain furnishes a rigorous proof for the magnon-crystal (MC) ground state, which is schematically shown in Fig. \ref{fig:ocbm} and mathematically given by the eigenvector  
\begin{eqnarray}
|{\rm MC}\rangle \!=\! \prod_{j=1}^N \! |\!\!\uparrow_{1,j}\rangle \!\otimes\! \frac{1}{2}
\Bigl(|\!\!\downarrow_{2,j}\uparrow_{3,j}\uparrow_{4,j}\uparrow_{5,j}\rangle 
\!&-&\!|\!\!\uparrow_{2,j}\downarrow_{3,j}\uparrow_{4,j}\uparrow_{5,j}\rangle  \nonumber \\
+|\!\!\uparrow_{2,j}\uparrow_{3,j}\downarrow_{4,j}\uparrow_{5,j}\rangle
\!&-&\!|\!\!\uparrow_{2,j}\uparrow_{3,j}\uparrow_{4,j}\downarrow_{5,j}\rangle\Bigr). \nonumber \\
\label{vecMC}  
\end{eqnarray}
The MC phase (\ref{vecMC}) has the character of a tensor-product state of the fully polarized monomeric spins and the bound one-magnon eigenstates, which trap a single magnon within each square plaquette due to a destructive quantum interference ensured by alternating signs of the probability amplitudes. Note furthermore that the MC ground state emerges in the highly frustrated parameter region $J_2>2J_1$ and sufficiently high magnetic fields that however do not exceed the saturation field $h<h_s=J_1+2J_2$.  

In the rest of the parameter space $J_2<2J_1$ the ground states of the spin-1/2 Heisenberg octahedral chain can be found by rewriting the zero-field part of the Hamiltonian (\ref{hamoc}) using the composite spin operator of the square plaquette $\hat{\bf{S}}_{c,j}=\hat{\bf{S}}_{2,j} + \hat{\bf{S}}_{3,j}+\hat{\bf{S}}_{4,j} + \hat{\bf{S}}_{5,j}$ and two composite operators for spin pairs from opposite corners of the square plaquette $\hat{\bf{S}}_{24,j}=\hat{\bf{S}}_{2,j} + \hat{\bf{S}}_{4,j}$ and $\hat{\bf{S}}_{35,j}=\hat{\bf{S}}_{3,j}+ \hat{\bf{S}}_{5,j}$
\begin{eqnarray}
\hat{\cal H} = J_1 \sum_{j=1}^{N} (\boldsymbol{\hat{S}}_{1,j} &+& \boldsymbol{\hat{S}}_{1,j+1}) \!\cdot\! \boldsymbol{\hat{S}}_{c, j} \nonumber \\
+ \frac{J_2}{2} \sum_{j=1}^{N} (\boldsymbol{\hat{S}}_{c, j}^2 &-& \boldsymbol{\hat{S}}_{24, j}^2  - \boldsymbol{\hat{S}}_{35, j}^2). 
\label{heffoc} 
\end{eqnarray}
The composite spin operators $\hat{\bf{S}}_{c,j}$, $\hat{\bf{S}}_{24,j}$ and $\hat{\bf{S}}_{35,j}$ commute with the Hamiltonian (\ref{heffoc}) and thus, they correspond to conserved quantities with well defined spin quantum numbers $S_{c,j} = 0$, 1 or 2; $S_{24,j} = 0$ or 1; $S_{35,j} = 0$ or 1. The spin-1/2 Heisenberg octahedral chain becomes equivalent to the effective mixed spin-(1/2, $S_c$) Heisenberg chains (see Fig. \ref{fig:eff}) whose eigenenergies are just trivially shifted due to the second term of the Hamiltonian (\ref{heffoc}) depending on the quantum spin numbers $S_{c,j}$, $S_{24,j}$ and $S_{35,j}$. 

\begin{figure}
 \centering\includegraphics[width=\columnwidth]{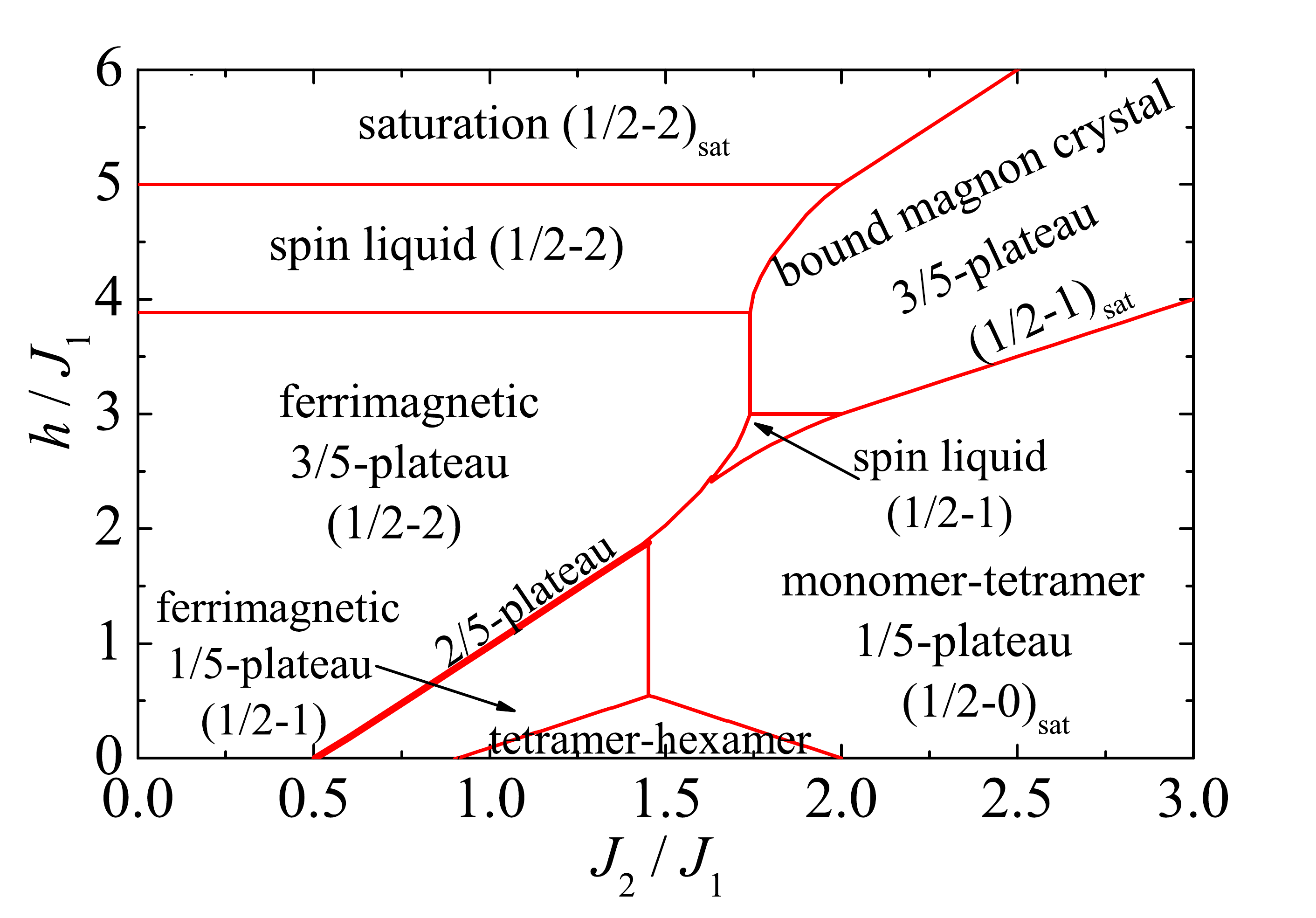} 
\vspace{-0.8cm}
\caption{The ground-state phase diagram of the spin-1/2 Heisenberg octahedral chain in $J_2/J_1-h/J_1$ plane.}
\label{fig:gsocw}
\end{figure}

The lowest-energy eigenstates of the effective mixed spin-(1/2, $S_c$) Heisenberg chains were previously obtained by extensive DMRG simulations \cite{stre17} implemented within the open-source ALPS software \cite{baue11}. If the four spins forming a square plaquette are in the singlet-tetramer state, i.e. the composite spin of a square plaquette becomes zero $S_{c,j} = 0$, then, the spin-1/2 Heisenberg octahedral chain is broken into smaller fragments due to a lack of spin-spin correlations across this square plaquette. A few exact fragmented ground states can be  found using this procedure. In this way one recovers for instance the MT ground state (\ref{vecMT}) by considering $\forall j$ $S_{c,j} = 0$  and the MC ground state (\ref{vecMC}) when assuming $\forall j$ $S_{c,j} = 1$. In addition, the regular alternation of the composite spin values $S_{c,2j-1} = 1$ and $S_{c,2j} = 0$ within the effective mixed spin-(1/2,1,1/2,0) Heisenberg chain results in the singlet tetramer-hexamer (TH) phase given by the eigenvector
\begin{widetext}
\begin{eqnarray}
|{\rm TH} \rangle = \prod_{j=1}^{N/2} \frac{1}{\sqrt{3}} 
\Bigl[|\!\uparrow_{2,2j-1}\downarrow_{3,2j-1}\uparrow_{4,2j-1}\downarrow_{5,2j-1}\rangle &+& 
|\!\downarrow_{2,2j-1}\uparrow_{3,2j-1}\downarrow_{4,2j-1}\uparrow_{5,2j-1}\rangle - \frac{1}{2} |\!\uparrow_{2,2j-1}\uparrow_{3,2j-1}\downarrow_{4,2j-1}\downarrow_{5,2j-1}\rangle \nonumber  \\
- \frac{1}{2} \Bigl(|\!\uparrow_{2,2j-1}\downarrow_{3,2j-1}\downarrow_{4,2j-1}\uparrow_{5,2j-1}\rangle &+& |\!\downarrow_{2,2j-1}\uparrow_{3,2j-1}\uparrow_{4,2j-1}\downarrow_{5,2j-1}\rangle 
+ |\!\downarrow_{2,2j-1}\downarrow_{3,2j-1}\uparrow_{4,2j-1}\uparrow_{5,2j-1}\rangle \Bigr) \Bigr] \nonumber  \\
\otimes \frac{1}{\sqrt{12}} \Bigl(
|\!\uparrow_{1,2j}\uparrow_{2,2j}\downarrow_{3,2j}\uparrow_{4,2j}\downarrow_{5,2j}\downarrow_{1,2j+1}\rangle 
&+& |\!\downarrow_{1,2j}\uparrow_{2,2j}\downarrow_{3,2j}\uparrow_{4,2j}\downarrow_{5,2j}\uparrow_{1,2j+1}\rangle 
+ |\!\uparrow_{1,2j}\downarrow_{2,2j}\uparrow_{3,2j}\downarrow_{4,2j}\downarrow_{5,2j}\uparrow_{1,2j+1}\rangle \nonumber \\ 
+ |\!\uparrow_{1,2j}\downarrow_{2,2j}\downarrow_{3,2j}\downarrow_{4,2j}\uparrow_{5,2j}\uparrow_{1,2j+1}\rangle 
&+& |\!\downarrow_{1,2j}\uparrow_{2,2j}\uparrow_{3,2j}\downarrow_{4,2j}\uparrow_{5,2j}\downarrow_{1,2j+1}\rangle 
+ |\!\downarrow_{1,2j}\downarrow_{2,2j}\uparrow_{3,2j}\uparrow_{4,2j}\uparrow_{5,2j}\downarrow_{1,2j+1}\rangle \nonumber \\
-|\!\uparrow_{1,2j}\downarrow_{2,2j}\uparrow_{3,2j}\downarrow_{4,2j}\uparrow_{5,2j}\downarrow_{1,2j+1}\rangle 
&-&|\!\downarrow_{1,2j}\downarrow_{2,2j}\uparrow_{3,2j}\downarrow_{4,2j}\uparrow_{5,2j}\uparrow_{1,2j+1}\rangle 
-|\!\uparrow_{1,2j}\uparrow_{2,2j}\downarrow_{3,2j}\downarrow_{4,2j}\downarrow_{5,2j}\uparrow_{1,2j+1}\rangle \nonumber \\
-|\!\uparrow_{1,2j}\downarrow_{2,2j}\downarrow_{3,2j}\uparrow_{4,2j}\downarrow_{5,2j}\uparrow_{1,2j+1}\rangle 
&-&|\!\downarrow_{1,2j}\uparrow_{2,2j}\uparrow_{3,2j}\uparrow_{4,2j}\downarrow_{5,2j}\downarrow_{1,2j+1}\rangle 
-|\!\downarrow_{1,2j}\uparrow_{2,2j}\downarrow_{3,2j}\uparrow_{4,2j}\uparrow_{5,2j}\downarrow_{1,2j+1}\rangle \Bigr). \nonumber \\
\label{vecTH}
\end{eqnarray}
\end{widetext}
The singlet TH phase is in fact two-fold degenerate, because another linearly independent eigenstate with the same energy can be obtained from the eigenvector (\ref{vecTH}) by interchanging singlet-tetramer and singlet-hexamer states on odd and even unit cells. The TH phase emerges just if the interaction ratio is from the interval $0.91<J_2/J_1<2.0$ and the magnetic field is sufficiently small $h/J_1<0.55$, whereby the highest possible field value for existence of the TH ground state is achieved for $J_2/J_1\approx 1.45$, see Fig.~\ref{fig:gsocw}.

\begin{figure}
\centering\includegraphics[width=0.9\columnwidth]{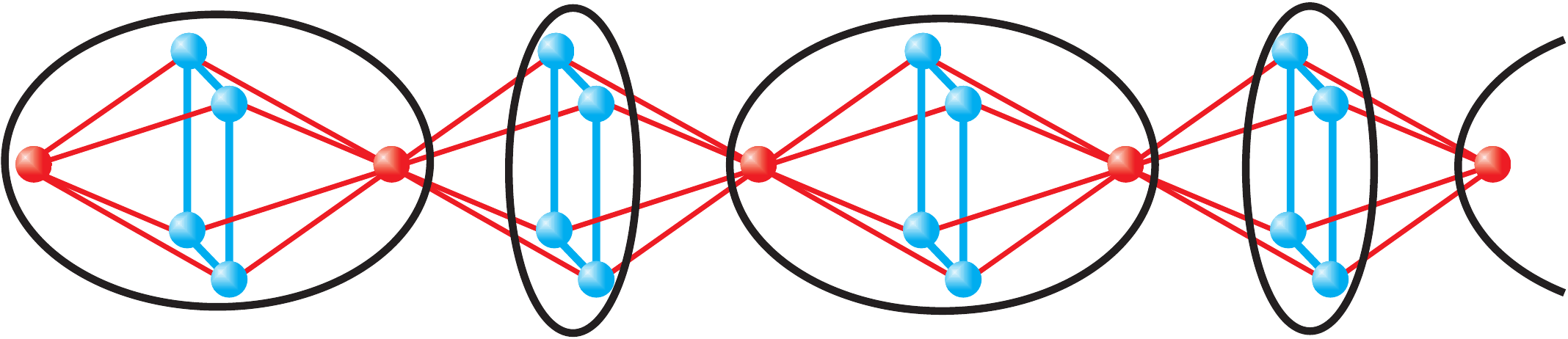} 
\vspace{-0.4cm}
\caption{A schematic illustration of the singlet tetramer-hexamer phase (\ref{vecTH}), which leads to occurrence of zero plateau in a zero-temperature magnetization curve. Small and large ovals correspond to the singlet-tetramer and singlet-hexamer state, respectively.}
\label{fig:octhw}
\end{figure} 

As can be seen in Fig. \ref{fig:gsocw}, the overall ground-state phase diagram of the spin-1/2 Heisenberg octahedral chain reveals a great diversity of remarkable quantum ground states including two ferrimagnetic Lieb-Mattis phases originating from the mixed spin-$(1/2,1)$ and mixed spin-$(1/2,2)$ Heisenberg chains, respectively, two quantum spin-liquid phases derived from the same couple of the mixed-spin Heisenberg chains closing an energy gap above the Lieb-Mattis ferrimagnetic phases, the gapful ferrimagnetic phase with a translationally broken symmetry, the fully polarized ferromagnetic phase and three additional fragmented ground states denoted as the MT phase [see Eq. (\ref{vecMT}) and Fig. \ref{fig:ocmt}], the bound MC phase [see Eq. (\ref{vecMC}) and Fig. \ref{fig:ocbm})] and the singlet TH phase [see Eq. (\ref{vecTH}) and Fig. \ref{fig:octhw}].

\section{Heisenberg octahedral chain as monomer-dimer problem}
\label{sec:ocmd}

In the following part we will investigate in detail low-temperature magnetization curves and thermodynamics of the spin-1/2 Heisenberg octahedral chain as extracted from the anticipated one-dimensional lattice-gas model of hard-core monomers and dimers. Recently, we have convincingly evidenced that the simpler version of the localized-magnon theory based on an effective one-dimensional lattice gas including two types of hard-core monomeric particles ascribed to bound one- and two-magnon eigenstates of square plaquettes provides within the highly frustrated parameter region $J_2/J_1>2$ a proper description of low-temperature magnetization curves and thermodynamics of the spin-1/2 Heisenberg octahedral chain from zero up to the saturation field \cite{stre17,stre18}. The main goal of the present work is to extend the validity of the localized-magnon theory to a less frustrated parameter space additionally involving the singlet TH phase (\ref{vecTH}). The idea and calculation procedure is quite similar as previously elucidated for the spin-1/2 Heisenberg diamond chain. 

\begin{figure}
\centering\includegraphics[width=0.9\columnwidth]{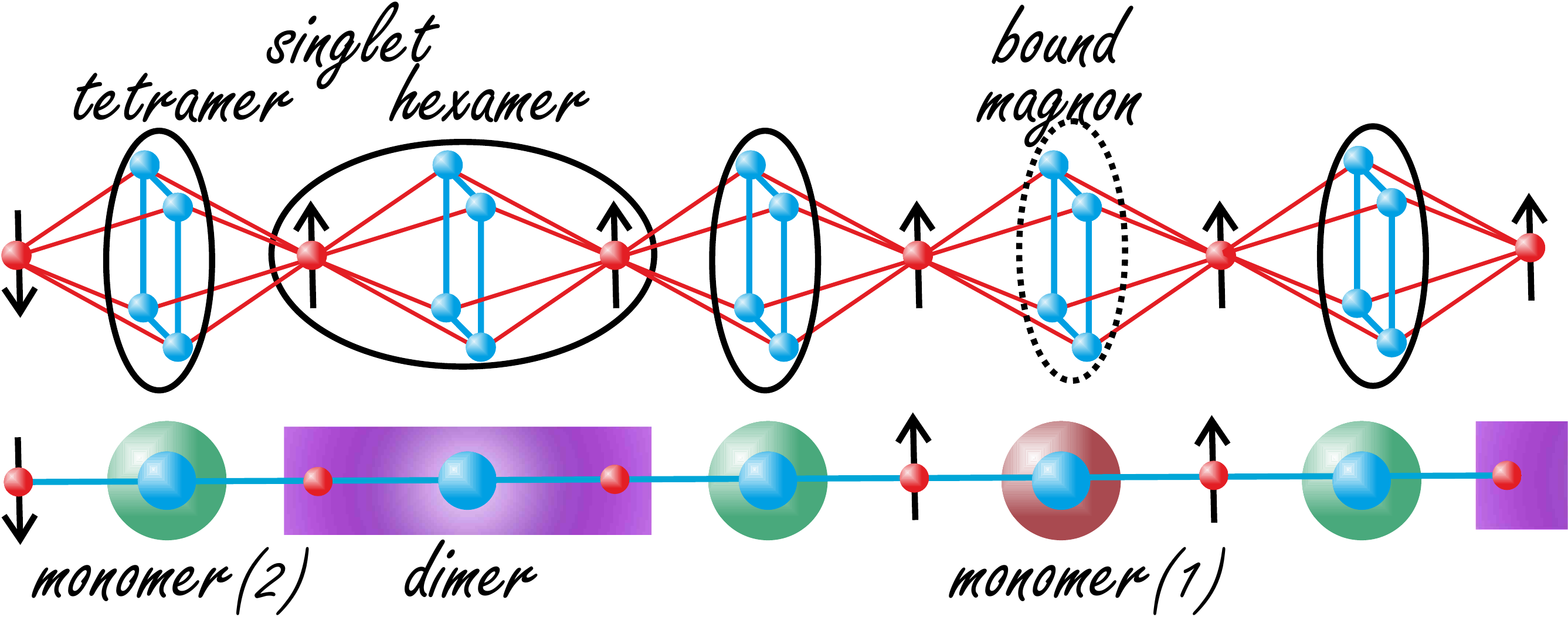}
\vspace{-0.4cm}
\caption{A schematic illustration of one paradigmatic eigenstate of the spin-1/2 Heisenberg octahedral chain and its equivalent representation within the three-component lattice-gas model of hard-core monomers and dimers. Small red spheres of the auxiliary lattice correspond to the monomeric spins $S_{1,j}$, while large blue spheres of the auxiliary lattice correspond to the composite spins $S_{c,j}$ assigned to the square plaquettes occupied by one of two monomeric particles or a dimeric particle.}
\label{auxLHC}
\end{figure}

The low-energy physics of the spin-1/2 Heisenberg octahedral chain will be effectively described by the lattice-gas model of hard-core monomers and dimers defined on auxiliary one-dimensional lattice, which may include the monomeric particle assigned to the singlet-tetramer (i.e. bound two-magnon) eigenstate of a square plaquette
\begin{eqnarray}
|m\rangle_j = 
  \frac{1}{\sqrt{3}} \Bigl[|\!\uparrow_{2,j}\downarrow_{3,j}\uparrow_{4,j}\downarrow_{5,j}\rangle &+& |\!\downarrow_{2,j}\uparrow_{3,j}\downarrow_{4,j}\uparrow_{5,j}\rangle \nonumber \\ 
	- \frac{1}{2} \Bigl(|\!\uparrow_{2,j}\uparrow_{3,j}\downarrow_{4,j}\downarrow_{5,j}\rangle &+& |\!\uparrow_{2,j}\downarrow_{3,j}\downarrow_{4,j}\uparrow_{5,j}\rangle \Bigr)  \label{sts} \\
  - \frac{1}{2} \Bigl(|\!\downarrow_{2,j}\uparrow_{3,j}\uparrow_{4,j}\downarrow_{5,j}\rangle &+& |\!\downarrow_{2,j}\downarrow_{3,j}\uparrow_{4,j}\uparrow_{5,j}\rangle \Bigr) \Bigr], \nonumber
\end{eqnarray}
the monomeric particle assigned to the bound one-magnon eigenstate of a square plaquette
\begin{eqnarray}
|n\rangle_j = \frac{1}{2} \Bigl(|\!\downarrow_{2,j}\uparrow_{3,j}\uparrow_{4,j}\uparrow_{5,j}\rangle 
&-& |\!\uparrow_{2,j}\downarrow_{3,j}\uparrow_{4,j}\uparrow_{5,j}\rangle \nonumber \\
+ |\!\uparrow_{2,j}\uparrow_{3,j}\downarrow_{4,j}\uparrow_{5,j}\rangle 
&-& |\!\uparrow_{2,j}\uparrow_{3,j}\uparrow_{4,j}\downarrow_{5,j}\rangle \Bigr),  
\label{bms}			
\end{eqnarray}
as well as, the dimeric particle assigned to the singlet-hexamer state of a octahedron (six-spin) cluster being a cornerstone of the TH ground states (\ref{vecTH})
\begin{eqnarray}
|d\rangle_j = \frac{1}{\sqrt{12}} \Bigl(
|\!\uparrow_{1,2j}\uparrow_{2,2j}\downarrow_{3,2j}\uparrow_{4,2j}\downarrow_{5,2j}\downarrow_{1,2j+1}  &\rangle& \nonumber \\
+ |\!\downarrow_{1,2j}\uparrow_{2,2j}\downarrow_{3,2j}\uparrow_{4,2j}\downarrow_{5,2j}\uparrow_{1,2j+1}&\rangle& \nonumber \\
+ |\!\uparrow_{1,2j}\downarrow_{2,2j}\uparrow_{3,2j}\downarrow_{4,2j}\downarrow_{5,2j}\uparrow_{1,2j+1}&\rangle& \nonumber \\
+ |\!\uparrow_{1,2j}\downarrow_{2,2j}\downarrow_{3,2j}\downarrow_{4,2j}\uparrow_{5,2j}\uparrow_{1,2j+1}&\rangle& \nonumber \\
+ |\!\downarrow_{1,2j}\uparrow_{2,2j}\uparrow_{3,2j}\downarrow_{4,2j}\uparrow_{5,2j}\downarrow_{1,2j+1}&\rangle& \nonumber \\
+ |\!\downarrow_{1,2j}\downarrow_{2,2j}\uparrow_{3,2j}\uparrow_{4,2j}\uparrow_{5,2j}\downarrow_{1,2j+1}&\rangle& \nonumber \\
- |\!\uparrow_{1,2j}\downarrow_{2,2j}\uparrow_{3,2j}\downarrow_{4,2j}\uparrow_{5,2j}\downarrow_{1,2j+1}&\rangle& \nonumber \\
- |\!\downarrow_{1,2j}\downarrow_{2,2j}\uparrow_{3,2j}\downarrow_{4,2j}\uparrow_{5,2j}\uparrow_{1,2j+1}&\rangle& \nonumber \\
- |\!\uparrow_{1,2j}\uparrow_{2,2j}\downarrow_{3,2j}\downarrow_{4,2j}\downarrow_{5,2j}\uparrow_{1,2j+1}&\rangle& \nonumber \\
- |\!\uparrow_{1,2j}\downarrow_{2,2j}\downarrow_{3,2j}\uparrow_{4,2j}\downarrow_{5,2j}\uparrow_{1,2j+1}&\rangle& \nonumber \\
- |\!\downarrow_{1,2j}\uparrow_{2,2j}\uparrow_{3,2j}\uparrow_{4,2j}\downarrow_{5,2j}\downarrow_{1,2j+1}&\rangle&\nonumber \\
- |\!\downarrow_{1,2j}\uparrow_{2,2j}\downarrow_{3,2j}\uparrow_{4,2j}\uparrow_{5,2j}\downarrow_{1,2j+1}&\rangle& \Bigr).
\label{shs}
\end{eqnarray}
One paradigmatic example of the feasible eigenstate of the spin-1/2 Heisenberg octahedral chain and its equivalent representation within the designed monomer-dimer lattice-gas model is illustrated in Fig. \ref{auxLHC}. Small red spheres of the auxiliary lattice correspond to lattice sites of the monomeric spins $S_{1,j}$ and large blue spheres of the auxiliary lattice correspond to lattice positions of the composite spins $S_{c,j} = S_{2,j}+S_{3,j}+S_{4,j}+S_{5,j}$ of the square plaquettes, which are available to the fully polarized ferromagnetic state acting as a reference vacuum state, the singlet-tetramer state (\ref{sts}) represented by the monomeric particle shown as a large green sphere, the bound one-magnon state (\ref{bms}) represented by the monomeric particle depicted as a large brown sphere, and the singlet-hexamer state (\ref{shs}) represented by the dimeric particle displayed as a large violet rectangle involving two neighboring monomeric spins $S_{1,j}$ and $S_{1,j+1}$ as well. In what follows, the energy of the square plaquettes is determined relative with respect to the fully polarized ferromagnetic state serving as a reference state and consequently, one should assign the chemical potential $\mu_1^{(2)} = 2J_1 + 3J_2 - 2h$ to the monomeric particles assigned to the bound two-magnon state (\ref{sts}), the chemical potential $\mu_1^{(1)} = J_1 + 2J_2 - h$ to the monomeric particles connected to the bound one-magnon state (\ref{bms}) and the chemical potential $\mu_2 = 3J_1 - h$ to the dimeric particles ascribed to the singlet-hexamer state (\ref{shs}). The effective monomer-dimer lattice-gas model can be then defined through the Hamiltonian 
\begin{eqnarray}
{\cal H} = E_{\rm FM}^0 - h \sum_{j=1}^N [2 &+& S_{1,j}^z (1-d_{j-1})(1-d_{j})] \nonumber \\
         - \mu_1^{(2)} \sum_{j=1}^{N} m_{j} &-& \mu_1^{(1)} \sum_{j=1}^{N} n_{j} - \mu_2 \sum_{j=1}^{N} d_{j}, 
\label{hefoc}				
\end{eqnarray}
where $E_{\rm FM}^0 \!=\! 2NJ_1 \!+\! N J_2$ is zero-field energy of the fully polarized ferromagnetic state, the second term is the Zeeman's term, $m_j\!=\!0,1$, $n_j\!=\!0,1$ and $d_j\!=\!0,1$ label occupation numbers for three kinds of particles additionally obeying hard-core constraint since each square plaquette of the spin-1/2 Heisenberg octahedral chain may be at most in one out of three considered eigenstates (\ref{sts})-(\ref{shs}). The partition function of the effective monomer-dimer lattice-gas model given by the Hamiltonian (\ref{hefoc}) can be calculated from the formula 
\begin{eqnarray}
{\cal Z} \!&=&\! {\rm e}^{-\beta E_{\rm FM}^0} \!\!\! \sum_{\{S_{1,j}^z\}} \! \sum_{\{m_{j}\}} \! \sum_{\{n_{j}\}} \! \sum_{\{d_{j}\}} \! \prod_{j=1}^N  \! 
\frac{1}{4^{d_j}} (1\!-\!d_{j-1} d_{j}) (1\!-\! m_j d_j) \nonumber \\
\!&\times&\! (1\!-\! n_j d_j) (1\!-\! m_j n_j)  \exp [ \beta (\mu_1^{(2)} m_{j} + \mu_1^{(1)} n_{j} + \mu_2 d_{j})] \nonumber \\
\!&\times&\! \exp [ \beta h S_{1,j}^z (1-d_{j-1})(1-d_{j})],
\label{pfefoc}
\end{eqnarray}
which already includes the factor $(1-d_{j-1} d_{j}) (1 - m_j d_j) (1\!-\! n_j d_j) (1\!-\! m_j n_j)$ ensuring a hard-core rule that prevents overlap and/or double occupancy of auxiliary lattice sites by the monomeric and dimeric particles. The other factor $\frac{1}{4^{d_j}}$ entering into the partition function (\ref{pfefoc}) is the correction term for the singlet-hexamer state (\ref{shs}), which avoids fourfold counting of this eigenstate when performing the summation over spin states of two monomeric spins $S_{1,j}$ and $S_{1,j+1}$. As in Sec.~\ref{sec:dcmd}, we note that the partition function above contains  the forbidden configurations when the singlet-plaquette state is followed by the fully polarized or bound one-magnon state on the neighboring dimer. But since the energy of these two configurations are rather high, they have no effect on the low-temperature thermodynamic properties studied here. After performing the summations $\sum_{\{S_{1,j}^z\}}$,  $\sum_{\{m_{j}\}}$ and $\sum_{\{n_{j}\}}$ over sets of the monomeric spins and monomeric hard-core particles one gets behind the product symbol the expression depending exclusively on the occupation numbers of the dimeric particles  
\begin{eqnarray}
\mbox{T} (d_{j}, d_{j+1}) \!&=&\! \frac{1}{4^{d_j}} (1\!-\!d_{j-1}d_{j}) \cosh \! \left[\frac{\beta h}{2} (1\!-\!d_{j-1})(1\!-\!d_j)\right] \nonumber \\ &\times&
2 {\rm e}^{2\beta h + \beta \mu_2 d_j} [1 + (1-d_j)({\rm e}^{\beta \mu_1^{(1)}}+{\rm e}^{\beta \mu_1^{(2)}})], \nonumber \\
\label{tmoc}
\end{eqnarray}
which can be repeatedly identified with the transfer matrix [cf. (\ref{dmtm})] further simplifying the exact calculation of the partition function according to the standard procedure
\begin{eqnarray}
{\cal Z} &=& {\rm e}^{-\beta E_{\rm FM}^0} \sum_{\{d_{j}\}} 
\prod_{j=1}^N  \mbox{T} (d_{j-1}, d_{j}) 
                     = {\rm e}^{-\beta E_{\rm FM}^0} \, \mbox{Tr} \, \mbox{T}^N   \nonumber \\
										&=& {\rm e}^{-\beta E_{\rm FM}^0} (\lambda_{+}^N + \lambda_{-}^N).
\label{pfoc}										
\end{eqnarray}
The final formula for the partition function (\ref{pfoc}) is expressed in terms of eigenvalues $\lambda_{\pm}$ of two-by-two transfer matrix (\ref{tmoc})  easily accessible through a direct diagonalization 
\begin{eqnarray}
\lambda_{\pm} &=& {\rm e}^{2\beta h}\Biggl\{ \! \cosh\!\left(\!\frac{\beta h}{2}\!\right) \!
\Xi_{o} \pm \sqrt{\left[\cosh\!\left(\!\frac{\beta h}{2}\!\right) \!
\Xi_{o} \right]^2 \!\!\!\! + {\rm e}^{\beta \mu_2} \, \Xi_{o}} \! \Biggr\}, \nonumber \\
\Xi_{o} &=& 1 + {\rm e}^{\beta \mu_1^{(1)}}+{\rm e}^{\beta \mu_1^{(2)}},
\label{chio}
\end{eqnarray}
which have quite analogous form to the one previously reported for the spin-1/2 Heisenberg diamond chain, cf. (\ref{chid}). 
The free-energy density of the finite-size spin-1/2 Heisenberg octahedral chain normalized per spin takes the following form
\begin{eqnarray}
f_{5N} &=& - k_{\rm B} T \frac{1}{5N} \ln {\cal Z}  \nonumber \\
&=& \frac{1}{5} \left(2J_1 + J_2 \right) - \frac{1}{5N} k_{\rm B} T \ln (\lambda_{+}^N + \lambda_{-}^N), 
\label{fedo}
\end{eqnarray}
which can be further simplified in the thermodynamic limit $N \to \infty$ when the free-energy density of the infinite spin-1/2 Heisenberg octahedral chain per spin is solely expressed in terms of the larger transfer-matrix eigenvalue
\begin{eqnarray}
f_{\infty} = - k_{\rm B} T \lim_{N \to \infty} \frac{1}{5N} \ln {\cal Z} = \frac{1}{5} \left(2J_1 \!+\! J_2 \right) \!-\! \frac{1}{5} k_{\rm B} T \ln \lambda_{+}. \nonumber \\
\label{fedon}
\end{eqnarray}
The final formulas (\ref{fedo}) and (\ref{fedon}) for the free-energy density may be subsequently employed for a straightforward calculation of the magnetization, susceptibility, entropy and specific heat of a finite and infinite spin-1/2 Heisenberg octahedral chain, respectively. The correctness of the effective description based on the monomer-dimer lattice-gas model (\ref{hefoc}) will be exemplified through a comparison of the as-obtained magnetization and specific-heat data with extensive numerical calculations employing the full ED and FTLM implemented within the open-source ALPS \cite{baue11} and Spinpack \cite{rich10,schu10} softwares. 

\begin{figure*}
  \includegraphics[width=0.49\textwidth]{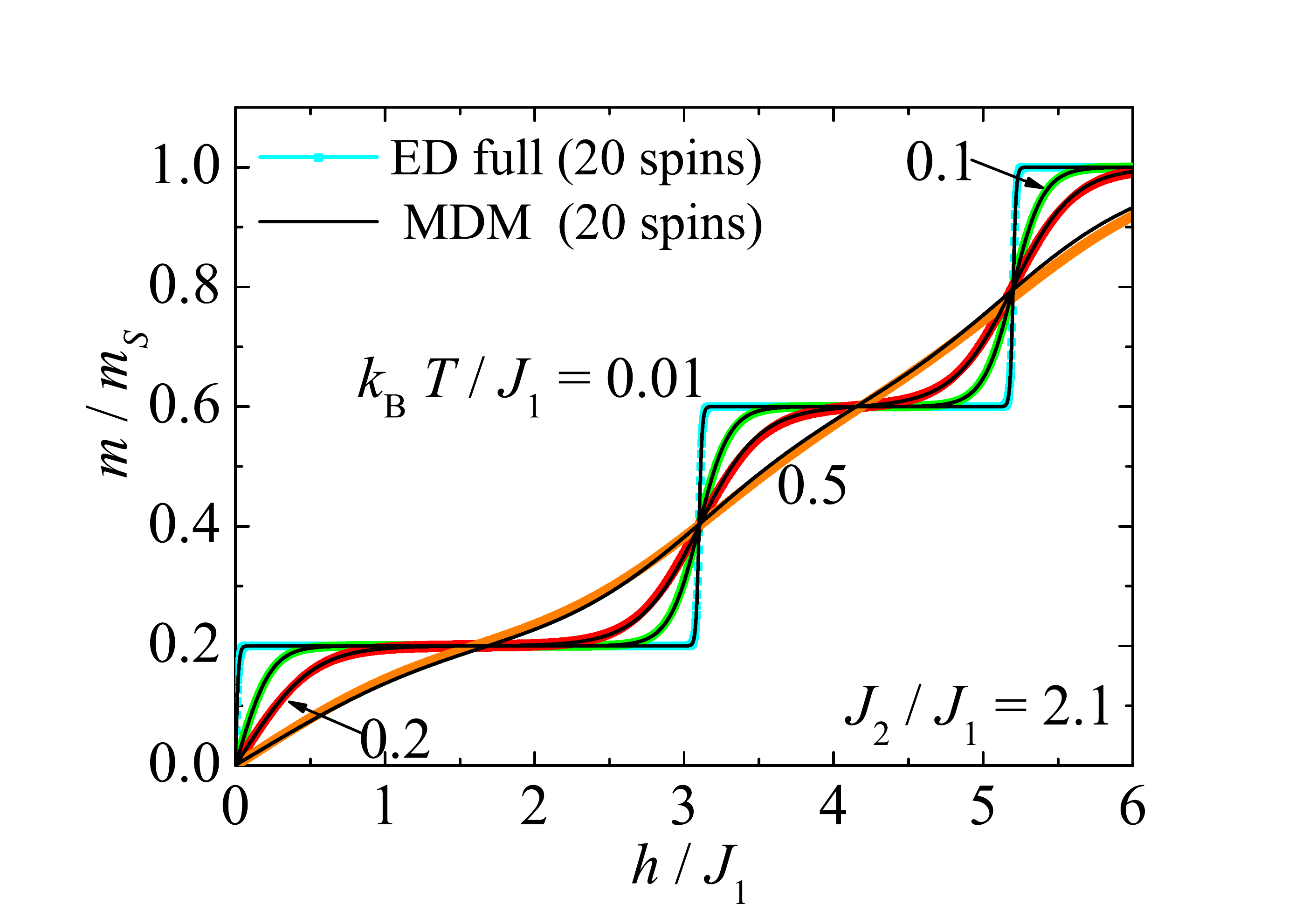}
  \hspace*{0.0cm}
  \includegraphics[width=0.49\textwidth]{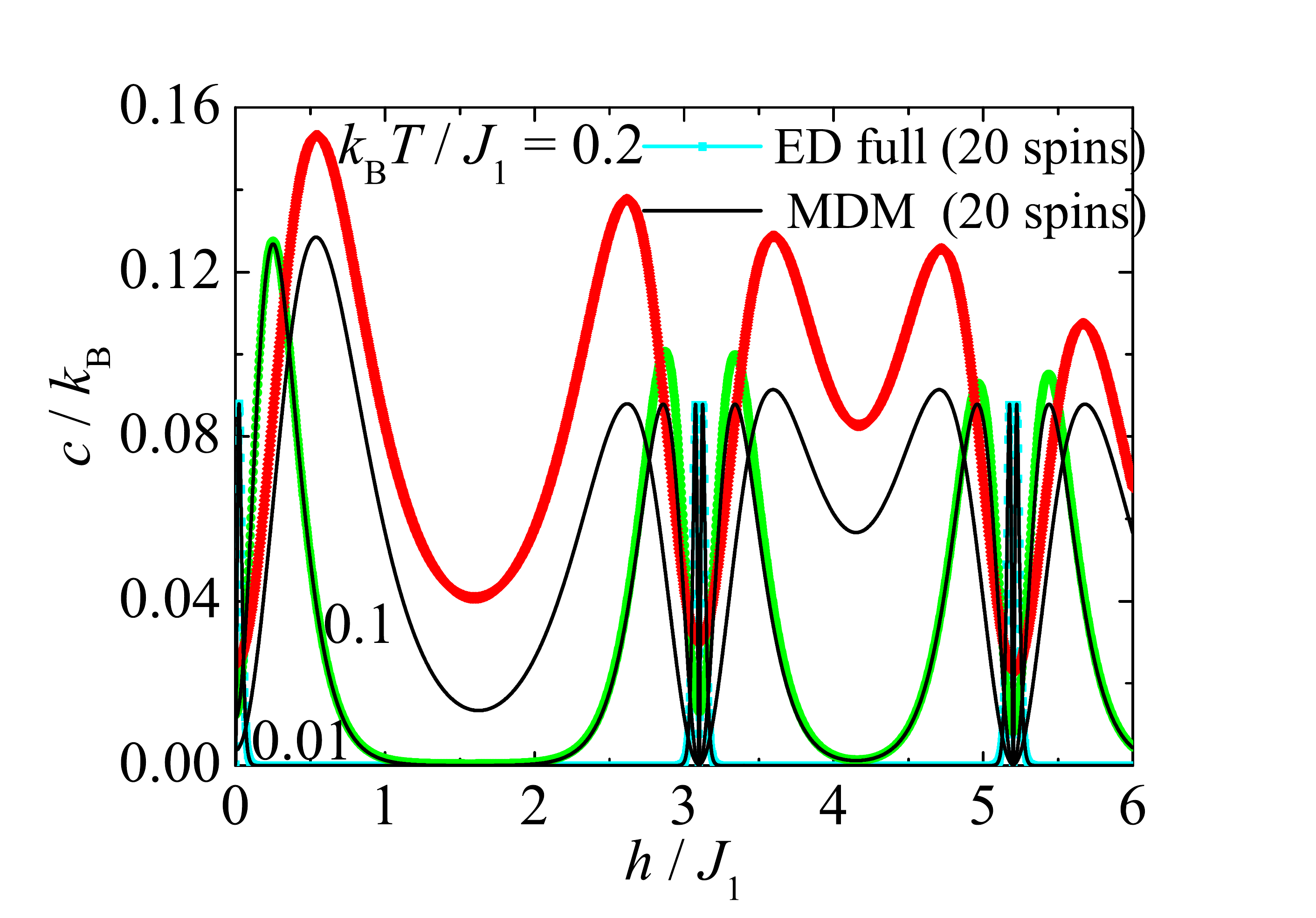} \\  
	\vspace*{-0.2cm}
  \includegraphics[width=0.49\textwidth]{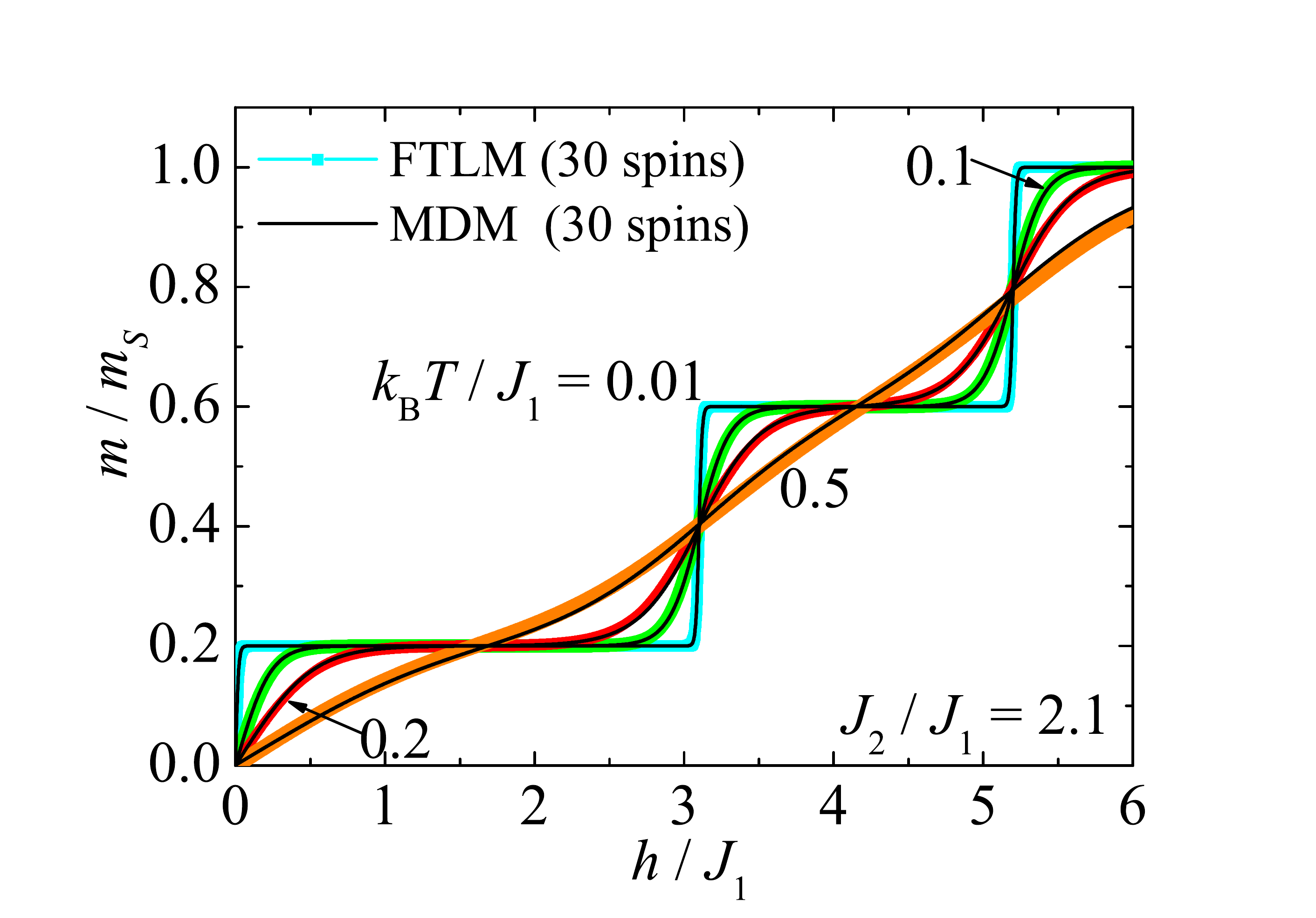}
  \hspace*{0.0cm}
  \includegraphics[width=0.49\textwidth]{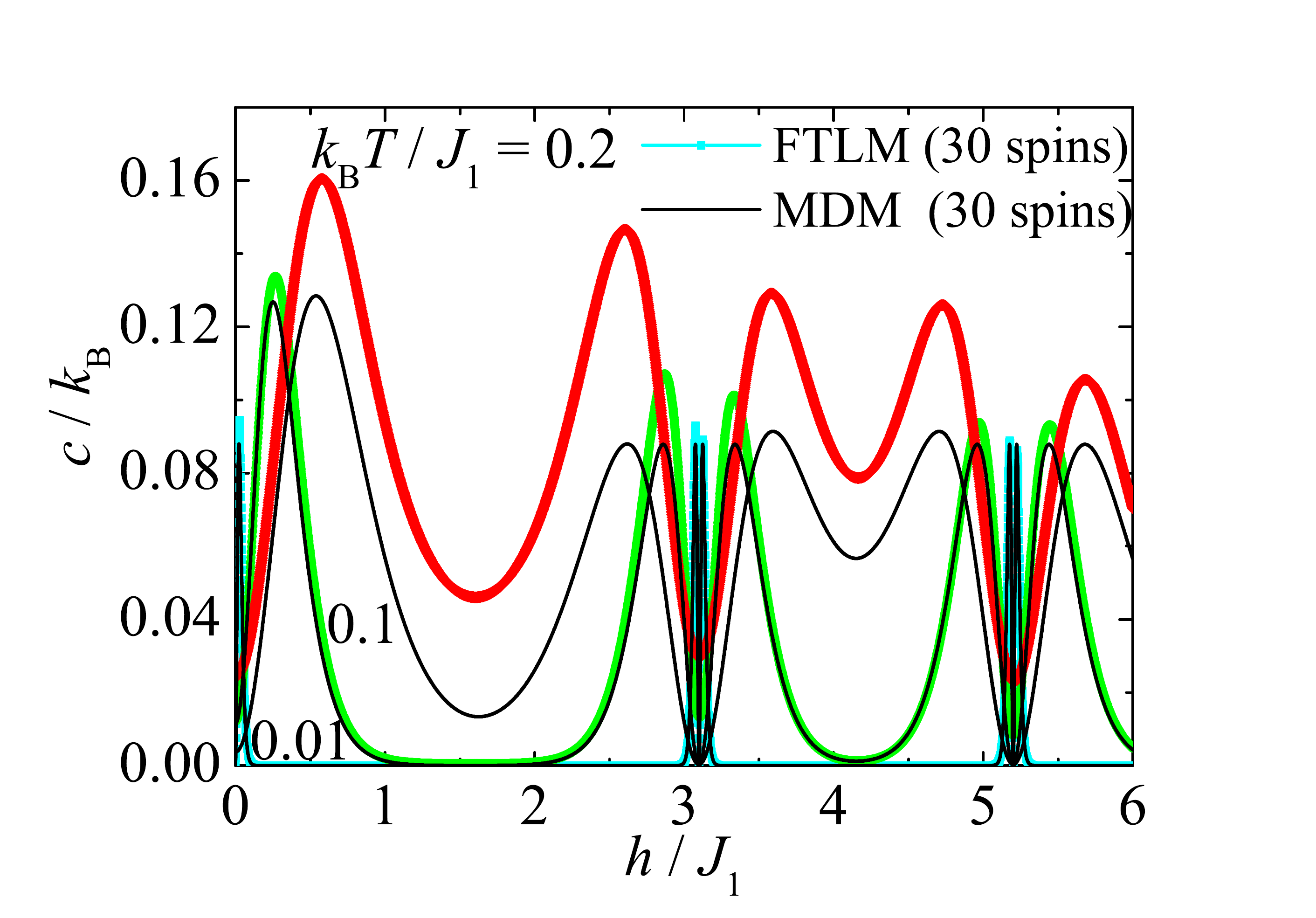}
	\vspace{-0.5cm}
	\caption{(Left panel) Magnetization curves of the spin-1/2 Heisenberg octahedral chain with $J_2/J_1=2.1$ obtained from the effective monomer-dimer model (MDM) versus the ED data for 20 spins (upper panel) and the FTLM data for 30 spins (lower panel) of the original model (\ref{hamoc}). (Right panel) The same as in the left panel but for the specific heat. Thin black lines show analytical results derived from the effective monomer-dimer model (\ref{hefoc}), while colored symbols refer to the ED (upper panel) or FTLM (lower panel) data.}
	\label{MPSHoc}
\end{figure*}
  
First, the magnetization and specific heat of the spin-1/2 Heisenberg octahedral chain are plotted in Fig. \ref{MPSHoc} as a function of the magnetic field for the interaction ratio $J_2/J_1=2.1$. The spin-1/2 Heisenberg octahedral chain apparently exhibits two intermediate magnetization plateaus at one-fifth and three-fifths of the saturation magnetization, which can be attributed to the MT and MC ground states unambiguously given by the eigenvectors (\ref{vecMT}) and (\ref{vecMC}), respectively. It directly follows from Fig. \ref{MPSHoc} that the magnetization curves descended from the effective monomer-dimer lattice-gas model are in a perfect agreement with the full ED data for 20 spins and FTLM data for 30 spins up to moderate temperature $k_{\rm B}T/J_1 \approx 0.5$. The perfect concordance is a direct consequence of a proper counting of localized many-magnon eigenstates (including among others two realized MT and MC ground states), which form a low-lying part of the energy spectrum being most relevant at low enough temperatures. On the other hand, the magnetic-field variations of the specific heat display more pronounced discrepancies already at much lower temperatures $k_{\rm B}T/J_1 \gtrsim 0.1$, above which collective quantum states not accounted in the anticipated monomer-dimer lattice-gas model are sufficiently thermally populated in order to cause conspicuous uprise with respect to the specific-heat value originating entirely from the many-magnon eigenstates. In spite of this shortcoming, the monomer-dimer lattice-gas model at least qualitatively reproduces a double-peak feature of the specific heat observable in a vicinity of all field-driven phase transitions at low and moderate temperatures. 
	
\begin{figure*}
  \includegraphics[width=0.49\textwidth]{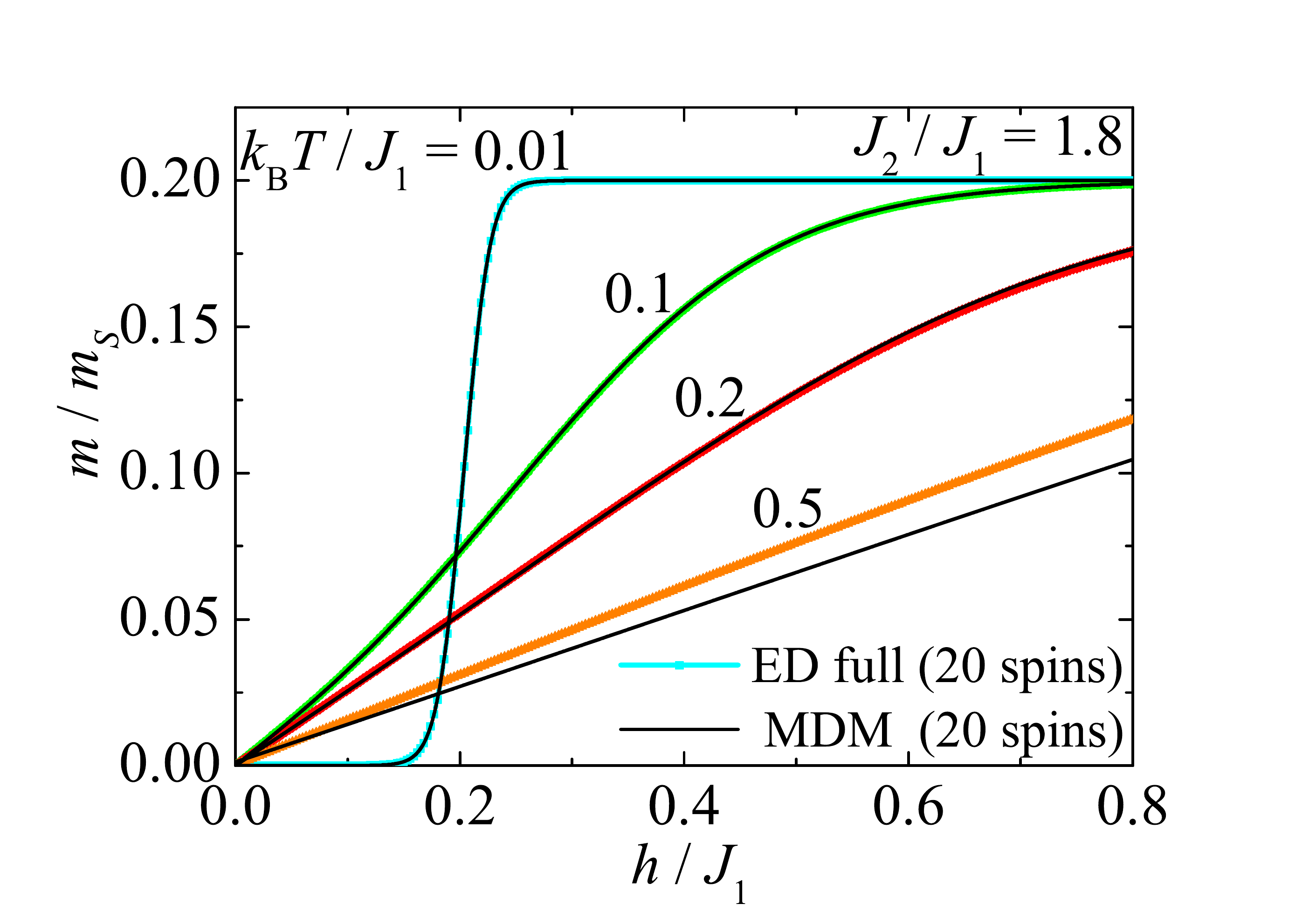}
  \hspace*{0.0cm}
  \includegraphics[width=0.49\textwidth]{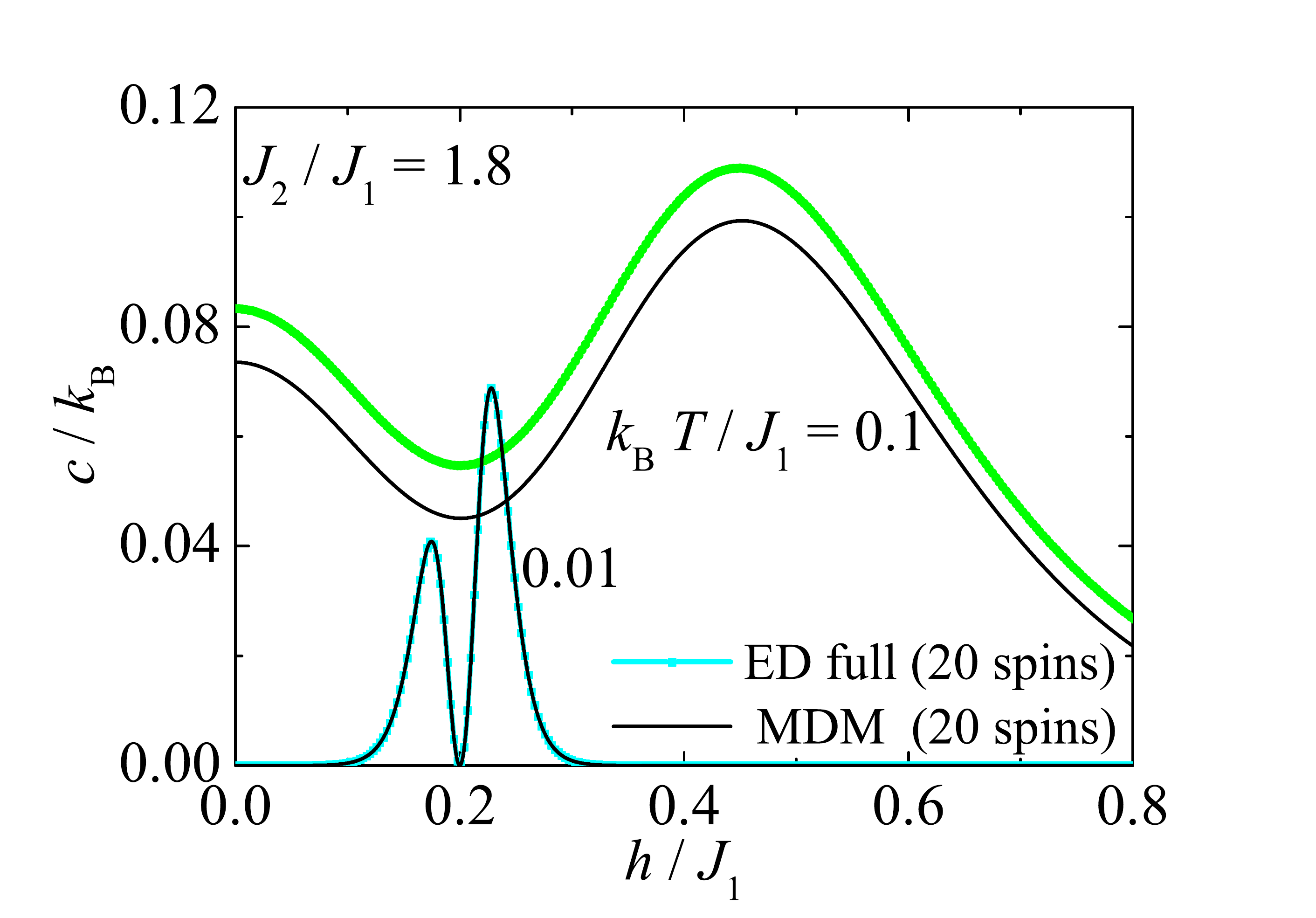} \\  
	\vspace*{-0.2cm}
  \includegraphics[width=0.49\textwidth]{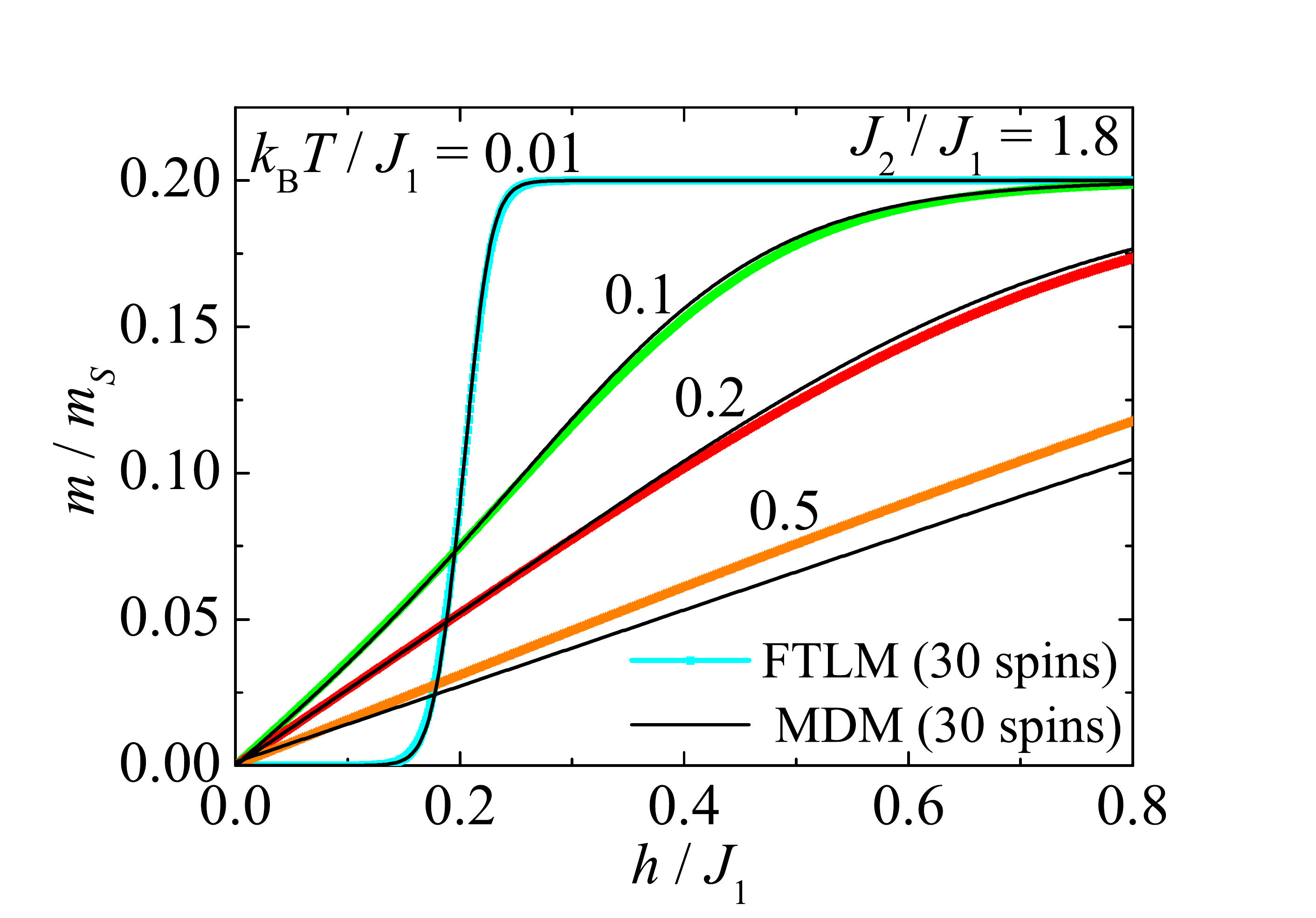}
  \hspace*{0.0cm}
  \includegraphics[width=0.49\textwidth]{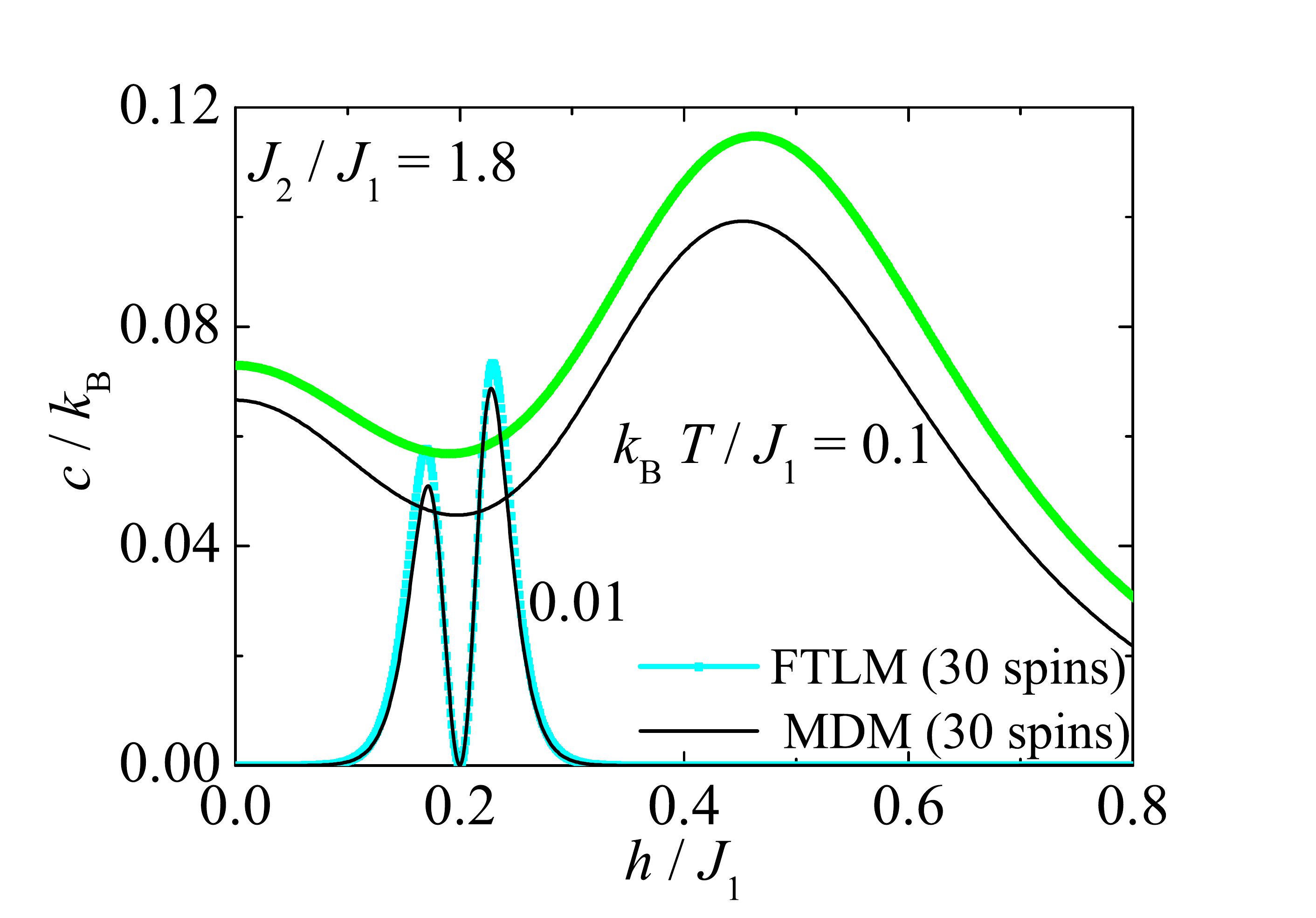}
	\vspace{-0.5cm}
	\caption{(Left panel) Magnetization curves of the spin-1/2 Heisenberg octahedral chain with $J_2/J_1=1.8$ obtained from the effective monomer-dimer model (MDM) versus the ED data for 20 spins (upper panel) and the FTLM data for 30 spins (lower panel) of the original model (\ref{hamoc}). (Right panel) The same as in the left panel but for the specific heat. Thin black lines show analytical results derived from the effective monomer-dimer model (\ref{hefoc}), while colored symbols refer to the ED (upper panel) or FTLM (lower panel) data.}
	\label{ocless}
\end{figure*}

Last but not least, the low-field part of magnetization curves of the spin-1/2 Heisenberg octahedral chain is depicted in the left panel of Fig. \ref{ocless} for the less frustrated parameter region $J_2/J_1=1.8$, where the investigated quantum spin chain undergoes a field-driven phase transition from the TH state (\ref{vecTH}) to the MT state (\ref{vecMT}). It can be understood from Fig. \ref{ocless} that the effective monomer-dimer model (\ref{hefoc}) not only properly reproduces the zero magnetization plateau attributable to the TH phase (\ref{vecTH}), but it also features the field-driven phase transition to the one-fifth plateau pertinent to the MT phase (\ref{vecMT}) and a gradual smoothing of the magnetization curves achieved upon rising the  temperature. The numerical ED and FTLM data indeed bear evidence to a legitimacy of this effective description up to moderate temperature $k_{\rm B}T/J_1 \approx 0.2$, which is however nearly a half of the temperature set as the upper bound for a reasonable description of the magnetization curves in the highly frustrated parameter space $J_2/J_1>2$. Furthermore, the specific heat of the spin-1/2 Heisenberg octahedral chain with the interaction ratio $J_2/J_1=1.8$ also displays an intriguing  double-peak dependence on the magnetic field as exemplified in the right panel of Fig. \ref{ocless}. The validity of the effective description of the specific heat of the spin-1/2 Heisenberg octahedral chain is again limited to relatively low temperatures $k_{\rm B}T/J_1 \lesssim 0.1$, above which the effective monomer-dimer model fails to reproduce the precise numerical ED and FTLM data at a quantitative level and provides qualitative insight only. Nevertheless, it should be  mentioned that the difference in the height of two round maxima of the specific heat around the field-induced phase transition can be simply interpreted in terms of the effective monomer-dimer lattice-gas model (\ref{hefoc}). The height of the specific-heat maximum is lower below the respective field-driven phase transition, because thermal excitations from the two-fold degenerate TH state (\ref{vecTH}) are less intense than the ones from the nondegenerate MT state (\ref{vecMT}).

\section{Conclusion}
\label{sec:con}

The present article is devoted to magnetic properties of the spin-1/2 Heisenberg diamond and octahedral chains, which are inferred from one-dimensional lattice-gas models of hard-core monomers and dimers deduced by means of the modified localized-magnon theory. An eligibility of the monomer-dimer lattice-gas models for elucidation of low-temperature magnetization curves and thermodynamic quantities of the spin-1/2 Heisenberg diamond and octahedral chains was decisively confirmed by state-of-the-art numerical calculations based on the ED and FTLM. Beside this, the ground-state phase diagrams of the spin-1/2 Heisenberg diamond and octahedral chains were established by making use of exact and DMRG calculations.

It has been evidenced that the anticipated monomer-dimer lattice-gas model provides a proper description of the spin-1/2 Heisenberg diamond chain when it is driven by the interaction parameters either to the singlet tetramer-dimer phase, the monomer-dimer phase or the saturated ferromagnetic phase. Similarly, the monomer-dimer lattice-gas model captures low-temperature features of the spin-1/2 Heisenberg octahedral chain provided that it is driven to the singlet tetramer-hexamer phase, the monomer-tetramer phase, the bound magnon-crystal phase or the saturated ferromagnetic phase. The developed localized-magnon theory also brings insight into a remarkable behavior of the specific heat, which shows a marked minimum at magnetic fields inherent to discontinuous field-driven phase transitions that is surrounded by two generally asymmetric round peaks originating from vigorous thermal excitations to low-lying localized many-magnon eigenstates.  Note furthermore that the previous localized-magnon theories \cite{derz06,stre17} valid only in a highly frustrated parameter region $J_2/J_1 > 2$ can be recovered from Eqs. (\ref{chid}) and (\ref{chio}) after neglecting the second term under the square root in the limit of sufficiently low temperatures and positive magnetic fields.  
						
Let us conclude our study by presenting a few future outlooks. We are convinced that the proposed calculation scheme has opened the route to a low-temperature magnetic behavior of several frustrated Heisenberg spin models, which have at least two tensor-product ground states with a magnetic unit spread over one and two unit cells, respectively. The mixed spin-(1,1/2) Heisenberg octahedral chain affords a suitable platform for implementation of the lattice-gas model of hard-core monomers and dimers, which would extend a legitimacy of the previous calculations based on the lattice-gas model of hard-core monomers \cite{karl19}. Note furthermore that the suggested calculation procedure is also compatible with recent extension enabling a straightforward computation of entanglement measures \cite{stre20}. Most importantly, the low-temperature physics of two-dimensional frustrated Heisenberg spin models with two tensor-product ground states, which have magnetic unit spread over one and two unit cells, would be captured by a two-dimensional lattice-gas model of hard-core monomers and dimers.
Finally, the spin-1/2 Heisenberg orthogonal-dimer chain \cite{schr02} and the mixed-spin Heisenberg diamond chain \cite{hida21} with an infinite series of fragmented quantum ground states provide another intriguing platform for further extension of the present approach, which would however require consideration of one-dimensional lattice-gas models of hard-core monomers, dimers, trimers, tetramers, etc.

\begin{acknowledgments}
J.Str. and K.K. acknowledge kind hospitality during summer 2021 at Max-Planck-Institut f\"ur Physik Komplexer Systeme in Dresden, where a part of this work was completed. J.Str. and K.K. acknowledge financial support provided by the grant The Ministry of Education, Science, Research and Sport of the Slovak Republic under the contract No. VEGA 1/0105/20 and by the grant of the Slovak Research and Development Agency under the contract No. APVV-20-0150. J.R. and J. Sch. acknowledge to DFG Grant.
\end{acknowledgments}

\end{document}